\documentclass[twocolumn,showpacs,amsfonts,aps,prc,nofootinbib,floatfix]{revtex4-1}

\usepackage{bm}
\usepackage{graphicx}
\usepackage{amsmath}
\usepackage{color}

\begin{document}

\title{Resonance Decay Contributions to Higher-Order Anisotropic Flow Coefficients}

\author{Zhi Qiu}
\author{Chun Shen}
\author{Ulrich Heinz}
\affiliation{Department of Physics, The Ohio State University,
  Columbus, OH 43210-1117, USA}

\begin{abstract}
We show that in hydrodynamic simulations for relativistic heavy-ion collisions, strong resonance decay calculations can be performed with fewer species of particle resonances while preserving good accuracy in single particle spectra and flow anisotropies. Such partial resonance calculations boost computation efficiency by a factor of 10 which is essential for large scale event-by-event simulations.
\end{abstract}


\date{\today}

\maketitle

\section{Introduction}
\label{sec:1}

Over the past decade, relativistic hydrodynamics has established itself as an indispensable component in modeling the collective dynamics of the quark-gluon plasma (QGP) produced in relativistic heavy ion collisions \cite{Kolb:2000sd,Teaney:2000cw,Huovinen:2001cy, Lacey:2006pn, Hirano:2005xf, Romatschke:2007mq, Luzum:2008cw, Luzum:2009sb, Song:2007ux, Song:2007fn, Song:2010mg, Luzum:2010ag, Lacey:2010ej, Bozek:2010wt, Hirano:2010jg, Schenke:2010rr, Schenke:2011tv, Song:2011qa, Shen:2011eg, Heinz:2011kt, Shen:2012vn}. The properties of the QGP, in particular its shear viscosity, can be deduced by studying its efficiency of converting initial geometric anisotropies into final anisotropic flows. Because of this, measurements of collective flow anisotropies $v_n$ (initially the elliptic flow $v_2$, more recently higher-order harmonic flows $v_{n\ge 3}$ \cite{Alver:2010gr, Adare:2011tg, Sorensen:2011fb, ALICE:2011vk, CMSflow, Steinberg:2011dj}) have generated strong interest and fertilized the study of initial geometry 
fluctuations and their dynamical consequences \cite{Holopainen:2010gz,Qin:2010pf,Qiu:2011iv,Dusling:2011rz,Flensburg:2011wx,Schenke:2011bn,Qiu:2011hf,Petersen:2012qc,Schenke:2012wb,Bozek:2012fw,Mota:2012zz,Chatterjee:2012dn,Jia:2012ma,Pang:2012he,Bozek:2012hy}.

The tool of choice for such studies is event-by-event hydrodynamics \cite{Holopainen:2010gz,Qin:2010pf,Qiu:2011iv,Schenke:2011bn,Qiu:2011hf,Petersen:2012qc,Schenke:2012wb,Bozek:2012fw,Mota:2012zz,Chatterjee:2012dn,Pang:2012he,Bozek:2012hy,Petersen:2010cw,Luzum:2010sp,Xu:2011fe,Luzum:2011mm, Schenke:2011bn, Shen:2011zc, Qiu:2011fi,Qiu:2012uy} where each fluctuating initial density profile is evolved separately, followed by taking an average over the event ensemble to compute $p_T$-spectra, anisotropic flow coefficients and two-particle correlations in the final state which can then be compared with experimental data. Due to the limited number of final state particles in each event, these observables can be measured with good statistical precision only as ensemble averages. Due to algorithmic progress over the last few years, the hydrodynamic evolution part is no longer the bottleneck in such event-by-event studies; at least for (2+1)-dimensional simulations (which assume longitudinal boost-invariance) the largest 
fraction of the computer time is spent converting the hydrodynamic output into final particle distributions, either on a ``switching surface" between a macroscopic hydrodynamic description of the QGP fluid and a microscopic kinetic evolution of the dilute late hadronic rescattering stage \cite{Hirano:2005xf,Song:2010aq}, or on a ``kinetic decoupling" surface marking the transition from a strongly coupled fluid directly to a non-interacting gas of free-streaming hadrons. The high numerical cost of this ``hydro-to-particle conversion" process results from the large number of unstable hadron resonances that need to be included and whose post-freeze-out decays (mostly due to strong-interaction processes, although for some comparisons with experimental data that have not been corrected \cite{Adams:2003xp,Adler:2003cb} for weak-decay feed-down, weak and electromagnetic decays must also be considered) modify the finally observed particle distributions.

The hydro-to-hadron conversion algorithm is based on the Cooper-Frye formula \cite{Cooper:1974mv} which expresses the final hadron momentum distribution as an integral of the local equilibrium (for ideal fluid dynamics) or slightly off-equilibrium (in viscous fluids) distribution function for the particle species in question over the conversion surface. Contributions to the spectra of experimentally measured {\em stable} particles from the strong decays of unstable resonances are then calculated from the single-particle spectra for the resonances \cite{Sollfrank:1990qz}. This requires the calculation of the directly emitted (``thermal'') particle momentum distributions for all $\sim300$ hadron species with mass typically up to 2 GeV
via Cooper-Frye integrals, followed by the evaluation of the phase-space integrals \cite{Sollfrank:1990qz} for all contributing decay channels. On a typical personal computer with a single-core CPU this calculation takes presently about 2-3 hours, compared to 10-15 minutes for the preceding hydrodynamic evolution with, say, the (2+1)-dimensional viscous fluid code {\tt VISH2{+}1} \cite{Song:2007fn,Song:2007ux,Song:2007ux}.

The $\sim2$\,GeV cutoff in resonance mass is dictated by requiring convergence of the relative particle yields of the measured hadronic final state after all unstable resonances have been allowed to decay. (The pion yields are especially sensitive to resonance feeddown.) Experimental evidence points to chemical decoupling at a temperature of $T_\mathrm{chem}\approx 165$\,MeV, i.e. close to the (pseudo)critical temperature for the quark-hadron phase transition \cite{BraunMunzinger:2003zd}; at this temperature, only resonances with masses above 2\,GeV are sufficiently strongly Boltzmann-suppressed that their decay contributions to stable particle yields can be safely ignored.

Here we show that for an accurate determination of the pion and proton anisotropic flow coefficients $v_n$ a much smaller number of resonances needs to be taken into account than for the hadron yields, and that even the shape of the azimuthally averaged pion and proton transverse momentum spectra can be reliably determined by accounting for only a small subset of the $\sim300$ resonance species mentioned above. These are the observables needed for an extraction of the QGP shear viscosity from heavy-ion collision experiments \cite{Song:2010mg}. By rearranging the resonance decay table in order of decreasing importance for the calculation of $p_T$-spectra and $v_n$ coefficients instead of increasing mass, good convergence for these observables can be achieved with a significantly reduced set of only about 20-30 resonances. This speeds up the computation by a factor of 10 -- a significant gain in efficiency for the iterative determination of the QGP shear viscosity.

The analysis presented here uses final states generated with the $(2{+1})$-dimensional boost-invariant viscous hydrodynamic code {\tt VISH2{+}1} for 200\,$A$\,GeV Au+Au collisions at the Relativistic Heavy-Ion Collider (RHIC) and for 2.76\,$A$\,TeV Pb+Pb collisions at the Large Hadron Collider (LHC) at various collision centralities, with previously determined \cite{Song:2010mg,Shen:2011eg,Qiu:2011hf} hydrodynamic input parameters. We find very similar results at both collision energies and therefore show here only plots for LHC collisions. Since the decay contributions from different resonances to the mentioned observables depend only on their decay channels and transverse momentum distributions, we expect little sensitivity to the assumption of longitudinal boost-invariance implicit in our approach and expect our reordered resonance decay tables to perform equally well for both $(2{+}1)$-d and $(3{+}1)$-d hydrodynamic simulations, and for a wide range of input parameters (such as QGP viscosity, 
thermalization time, initial entropy and energy density, etc.).

\section{Resonance ordering}
\label{sec:2}

The momentum distributions of directly emitted (``thermal'') resonances of species $i$ are computed from the Cooper-Frye formula \cite{Cooper:1974mv}:
\begin{equation} \label{eq:1}
  E \frac{dN_i}{d^3p} = \frac{dN}{dy\,p_T\,dp_T\,d\phi_p} =
  \frac{g_i}{(2 \pi)^3} \int_\Sigma \,p^\mu d^3\sigma_\mu (f_{i0} + \delta f_i).
\end{equation}
Here $\Sigma$ is the hydro-to-hadron conversion hypersurface, $d^3\sigma_\mu$ its surface normal vector, $f_{i0}=1/[e^{\beta (p \cdot u - \mu_i)} \mp 1]$ is the Bose or Fermi thermal equilibrium distribution function, and $\delta f_i$ accounts for viscous corrections (driven by the viscous pressure tensor $\pi_{\mu\nu}(x)$ on the conversion surface) of the local phase-space distribution along $\Sigma$. We assume the quadratic form $\delta f =\frac{1}{2} f_0 (1\pm f_0) \frac{p^\mu p^\nu}{T^2}\frac{ \pi_{\mu\nu}}{(e+p)}$.

Resonance decays increase the total yields of the stable hadrons and change their momentum distributions. For kinematic reasons, most of the light decay daughters have low transverse momenta, thus modifying the shape of light stable hadrons (pions, kaons) particle spectra mostly in the region $p_T<1.5$\,GeV \cite{Sollfrank:1990qz}. We denote the total decay contribution to the momentum distribution of stable hadron species $i$ by $\delta\!\left(dN_i/(dy d^2p_T)\right)$, and the total spectrum (obtained by adding this to the thermally emitted spectrum $dN_i^\mathrm{th}/(dy d^2p_T)$) by $dN_i^\mathrm{tot}/(dy d^2p_T)$. (We here include only strong and electromagnetic decays.) The $\bm{p}_T$-integrated total yield $\delta(dN_i/dy)$ of decay products of species $i$ is denoted by $\delta N_i$, with $N_i^\mathrm{tot}{\,=\,}N_i^\mathrm{th}{+}\delta N_i{\,=\,}N_i^\mathrm{th}{+}\sum_j \tilde{b}_{j\to i} N_j^\mathrm{th}$, where the sum is over resonances $j$ and $\tilde{b}_{j\to i}$ is the effective branching ratio (see Eq.~(\ref{eq4}) below) for the decay $j{\,\to\,}i$.

The contribution to $\delta N_i$ from a particular resonance $j$ is not only influenced by its mass (through the Boltzmann suppression factor $\sim e^{-E_j/T}$), but also by its spin degeneracy factor $g_j$ and its branching ratio $\tilde{b}_{j\to i}$ into the decay channel that feeds stable particle species $i$. For each stable hadron species $i$ it is therefore a different set of resonances that makes the most important contributions. Our goal is to order the resonances in decreasing order of importance, for each stable particle species $i$. We here assume that the conversion surface has constant temperature $T_\mathrm{conv}$. The different hadron resonances have $T_\mathrm{conv}$-dependent non-equilibrium fugacities $\lambda_j$ that ensure constant stable particle ratios equal to their chemical equilibrium values at $T_\mathrm{chem}$ and $\mu_B{\,=\,}0$, independent of the hydro-to-hadron conversion temperature $T_\mathrm{conv}$. While the actual fractions contributed by each resonance to the stable particle 
yields depend on $T_\mathrm{conv}$, the ordering of these fractions is largely $T_\mathrm{conv}$-independent.

We start from the resonance table in the {\tt AZHYDRO} package,\footnote{{\tt AZHYDRO} is available at the URL\\{\tt http://www.physics.ohio-state.edu/\~{}froderma/}.} which includes  $319$ species of hadrons (counting different isospin states such as $\pi^+$, $\pi^0$, $\pi^-$ as separate species) with rest masses up to $2.25$ GeV. After fixing the value of $T_\mathrm{conv}$ we look up the non-equilibrium fugacity $\lambda_j$ for each of these 319 species
from the EOS s95p-PCE tables constructed in Ref.~\cite{Huovinen:2009yb}. For each stable particle species $i$ we then generate an ordered list of resonances $j$ that can decay directly into $i$.
Note that in this ordering we account not only for direct decay contribution but also for multi-step decay cascades, where $j$ first decays into an unstable resonance $k$ which further decays (directly or through more intermediate steps) into the stable species $i$.

Table \ref{tab:1} shows the beginning of this contribution table for positively charged pions, for a conversion temperature $T_\mathrm{conv}{\,=\,}120$\,MeV. The ``total contribution'' percentages $c_{j\to i}$ in the third column are computed as
\begin{eqnarray}
  c_{j\to i}&=&\frac{N_i^{(j)}}{\sum_{j'}N_i^{(j')}}
                  =  \frac{\tilde{b}_{j{\to}i}N_j^\mathrm{th}}
                            {\sum_{j'} \tilde{b}_{j'\to i} N_{j'}^\mathrm{th}}, 
\label{eq2}
\\
 N_{j}^\mathrm{th} &=&
 g_j m_j^2 \sum_{k=1}^\infty \frac{(\pm)^{k+1}}{k} \lambda_j^k K_2\left(k\frac{m_j}{T}\right),
 \label{eq3}
\end{eqnarray}
where the effective branching ratios $\tilde{b}_{j{\to}i}$ in Eq.~(\ref{eq2}) account for multi-step decay cascades as follows:
\begin{eqnarray}
\label{eq4}
     \tilde{b}_{j{\to}i} =  b_{j\to i} &+& \sum_{k_1} b_{j\to k_1} b_{k_1\to i} 
     \nonumber\\
   &+& \sum_{k_1,k_2} b_{j\to k_1} b_{k_1\to k_2} b_{k_2\to i} + \dots\, .
\end{eqnarray}
%
%
\begin{figure*}[ht]
  \center{
	  \includegraphics[width=8cm,height=7cm]{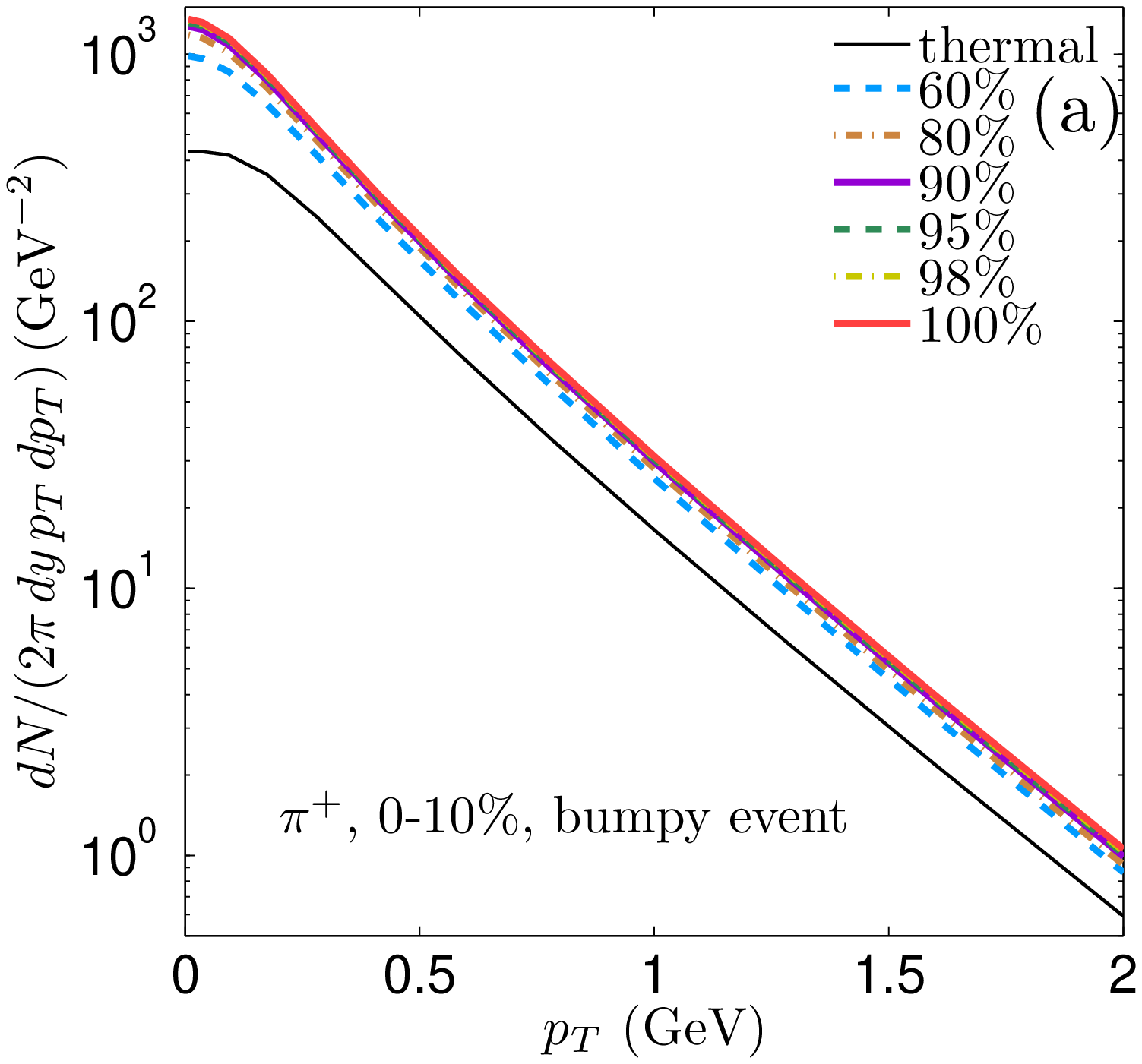}
	  \includegraphics[width=8cm,height=7cm]{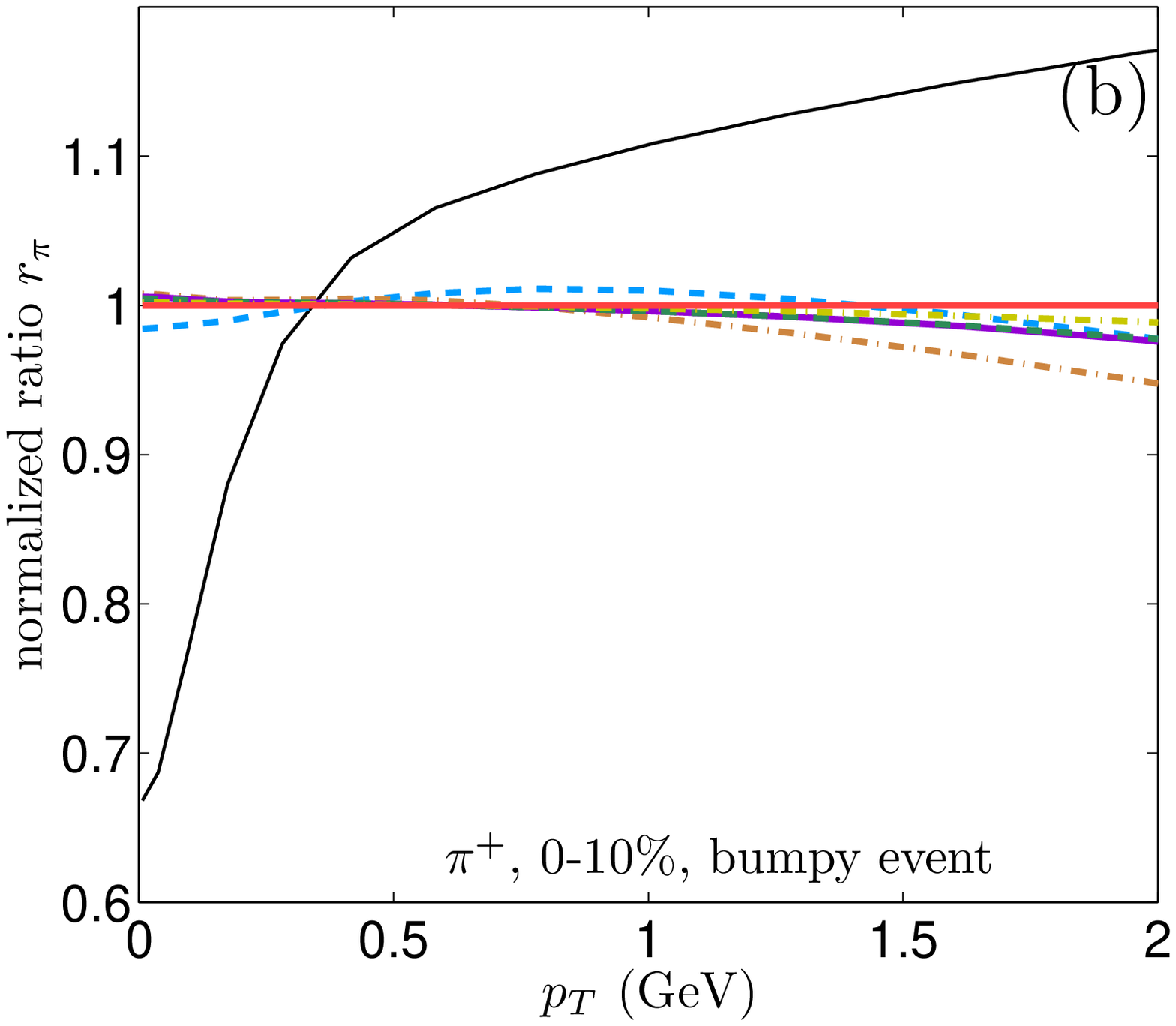}\\
	  \includegraphics[width=8cm,height=7cm]{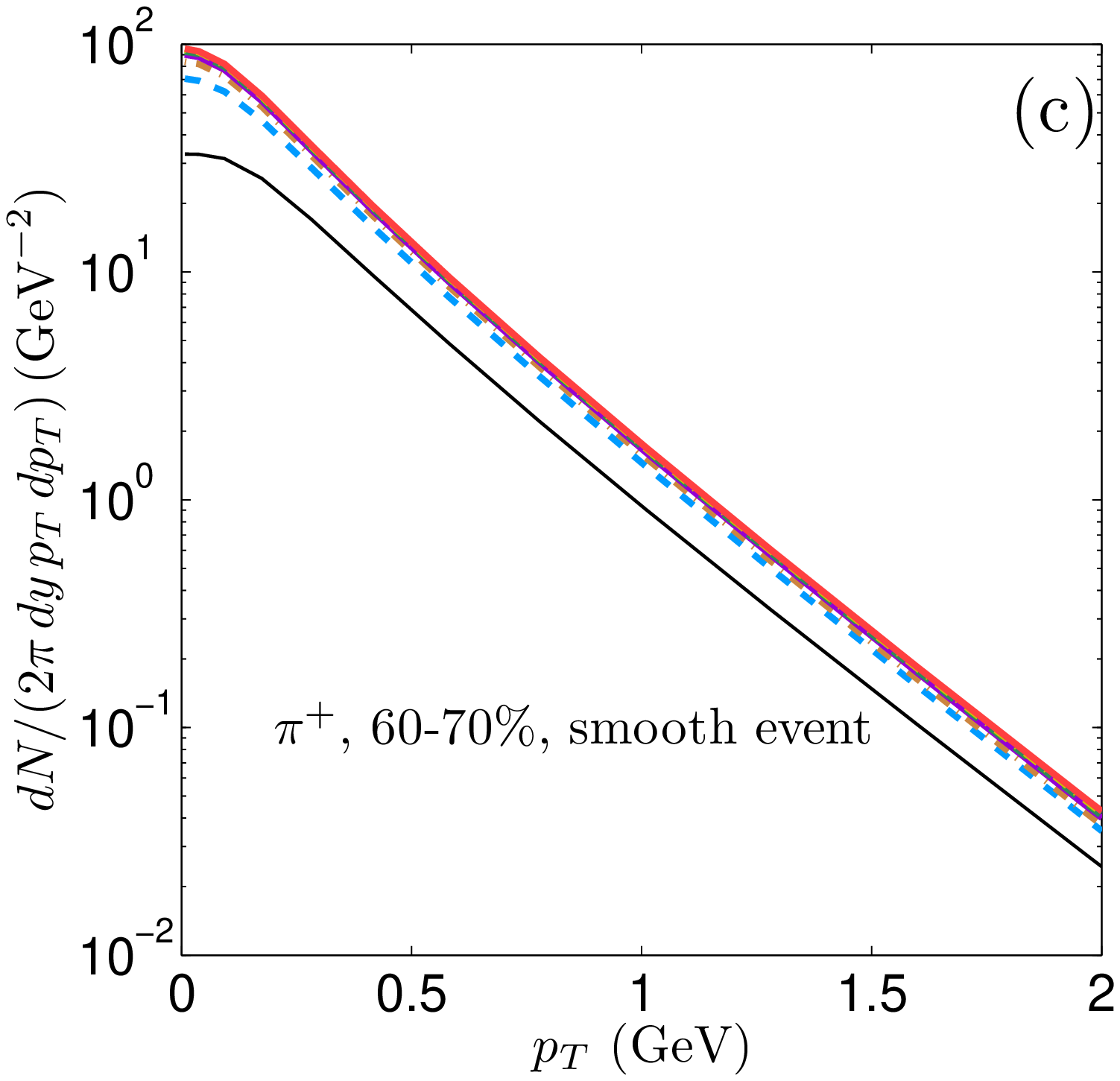}
	  \includegraphics[width=8cm,height=7cm]{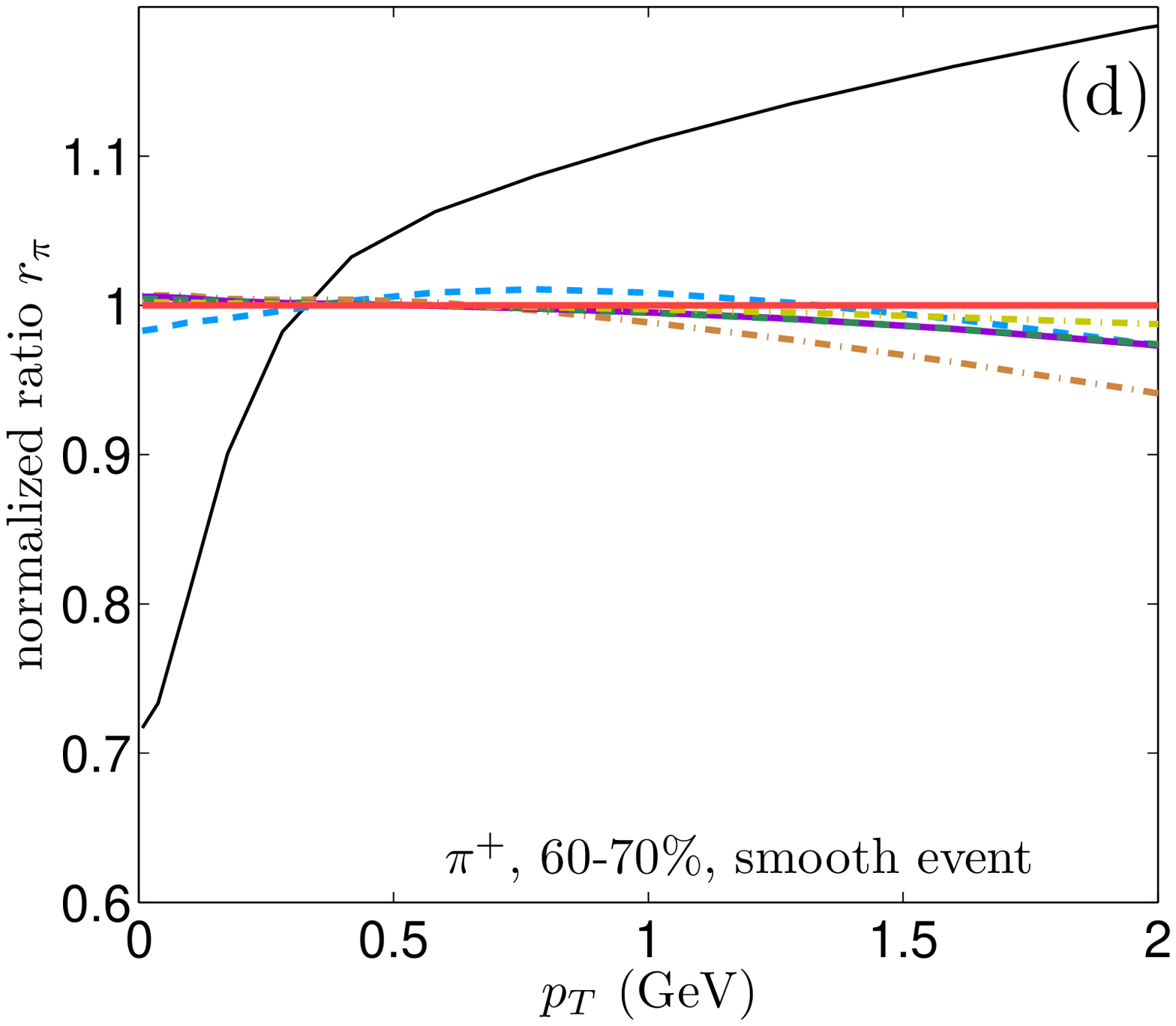}
  }
  \caption{(Color online) Transverse momentum spectra of $\pi^+$, for a bumpy central 
  (0-10\% centrality, top panels) and a smooth peripheral (60-70\% centrality, bottom panels)
  Pb-Pb collision at LHC energies. 
  Panels (a) and (c) present the absolutely normalized spectra, while panels (b) and (d) show 
  the normalized ratio $r_\pi(p_T)$ defined in Eq.~(\ref{ratio}). Different lines correspond 
  to different cumulative resonance decay contributions between 0\% (``thermal'') and 100\%.
  See text for discussion. 
\label{fig:1}}
\end{figure*}
%
\begin{table}
\begin{tabular}{|c|c|c|}
	\hline
	name & mass (GeV) & total contribution (\%) \\ \hline
	$\omega$ & 0.78260 & 15.398  \\ \hline
	$\rho^0$ & 0.77580 & 11.179  \\ \hline
	$\rho^+$ & 0.77580 & 11.098  \\ \hline
	...... & ...... & ......
\end{tabular}
\caption{Example of the $\pi^+$ contribution table for $T_\mathrm{conv}{\,=\,}120$\,MeV.}
\label{tab:1}
\end{table}
%
The sum over $k$ in (\ref{eq3}) takes care of quantum statistical effects, with the upper (lower) sign for bosons (fermions). For all hadrons except pions accurate results can be obtained by keeping 
only the first term $k{\,=\,}1$, i.e. by ignoring quantum statistical effects. Even for pions, a few $k$-terms suffice for good precision (in our calculations we truncate the series in (\ref{eq3}) at $k{\,=\,}10$). The complete ordered resonance decay contribution tables for $\pi^+$, $K^+$, $p$, $\Lambda$, $\Sigma^+$ and $\Xi^-$ are given in the Appendix. Horizontal lines in the tables indicate where the cumulative resonance decay contributions $c_i^\mathrm{cut}=\sum_{j{=}1}^{j_\mathrm{cut}} c_{j\to i}$ exceed certain threshold percentages (as indicated) of the total resonance decay contribution to species $i$.

In the following section we show the stable hadron $p_T$-spectra and their anisotropic flow coefficients as a function of these cumulative decay contribution percentages $c_i^\mathrm{cut}$, in order to assess how many resonances from these ordered decay tables should be included for an accurate computation of these observables.

%
\begin{figure*}[ht]
  \center{
	  \includegraphics[width=8cm,height=7cm]{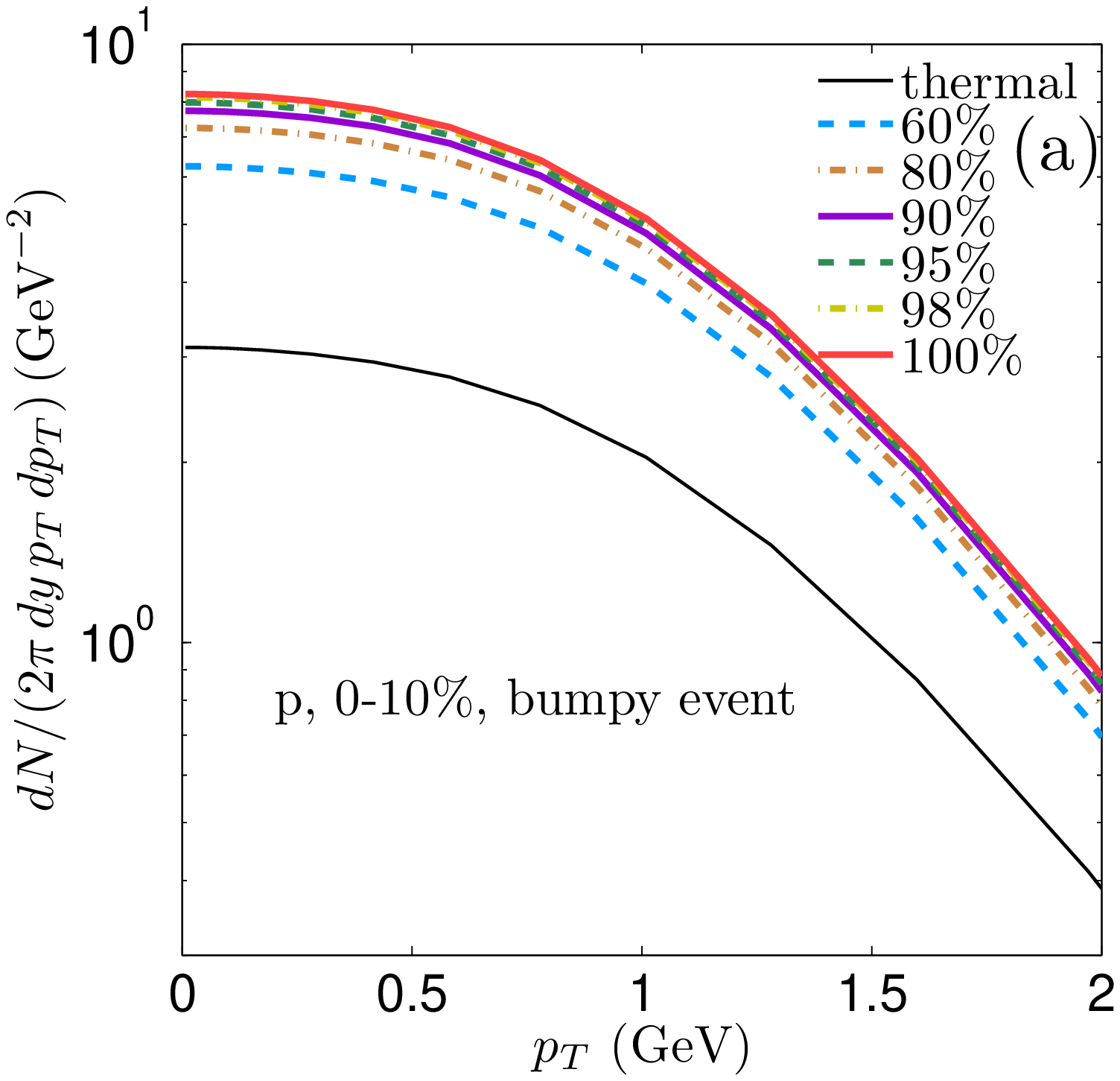}
	  \includegraphics[width=8cm,height=7cm]{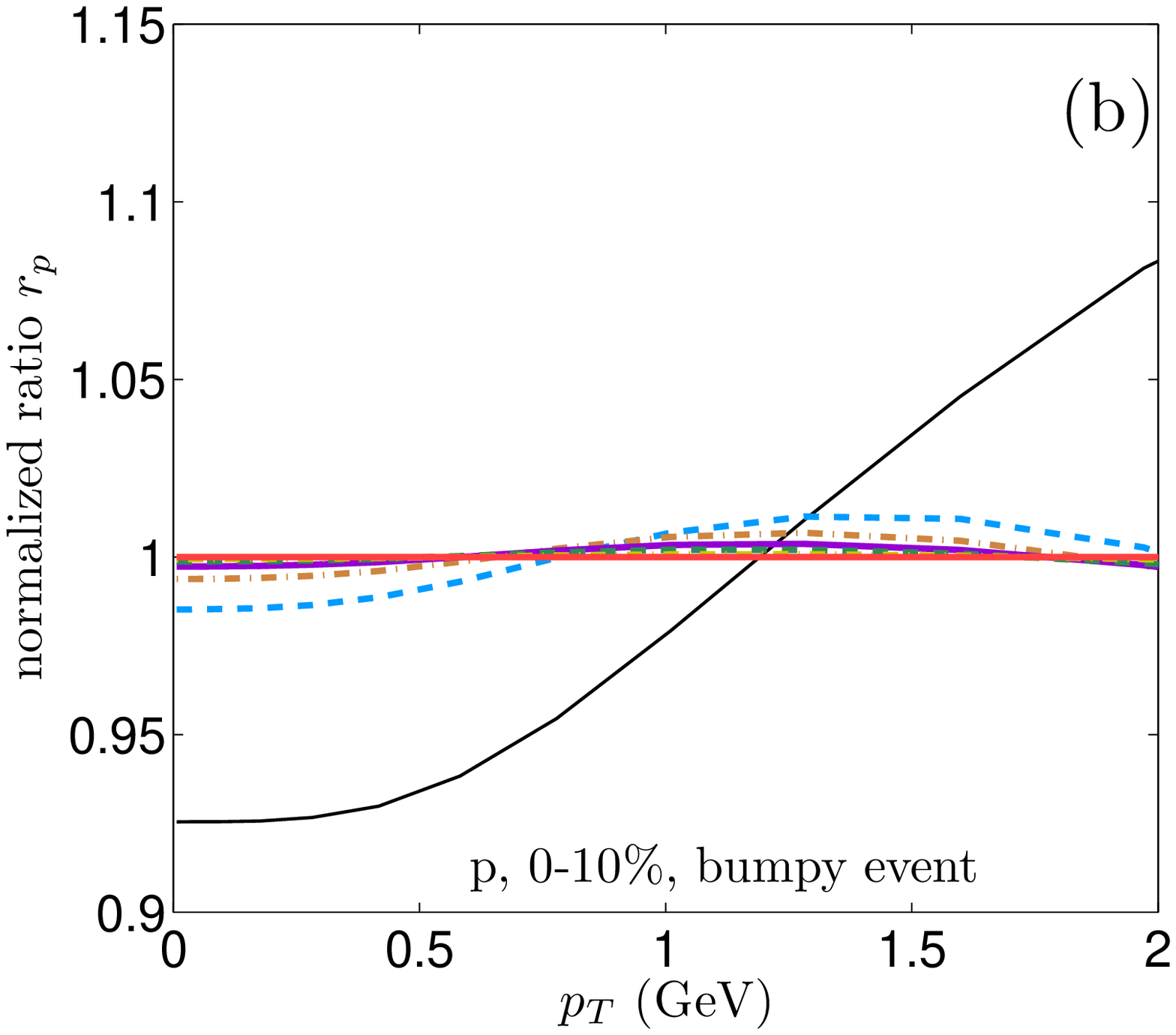}\\
	  \includegraphics[width=8cm,height=7cm]{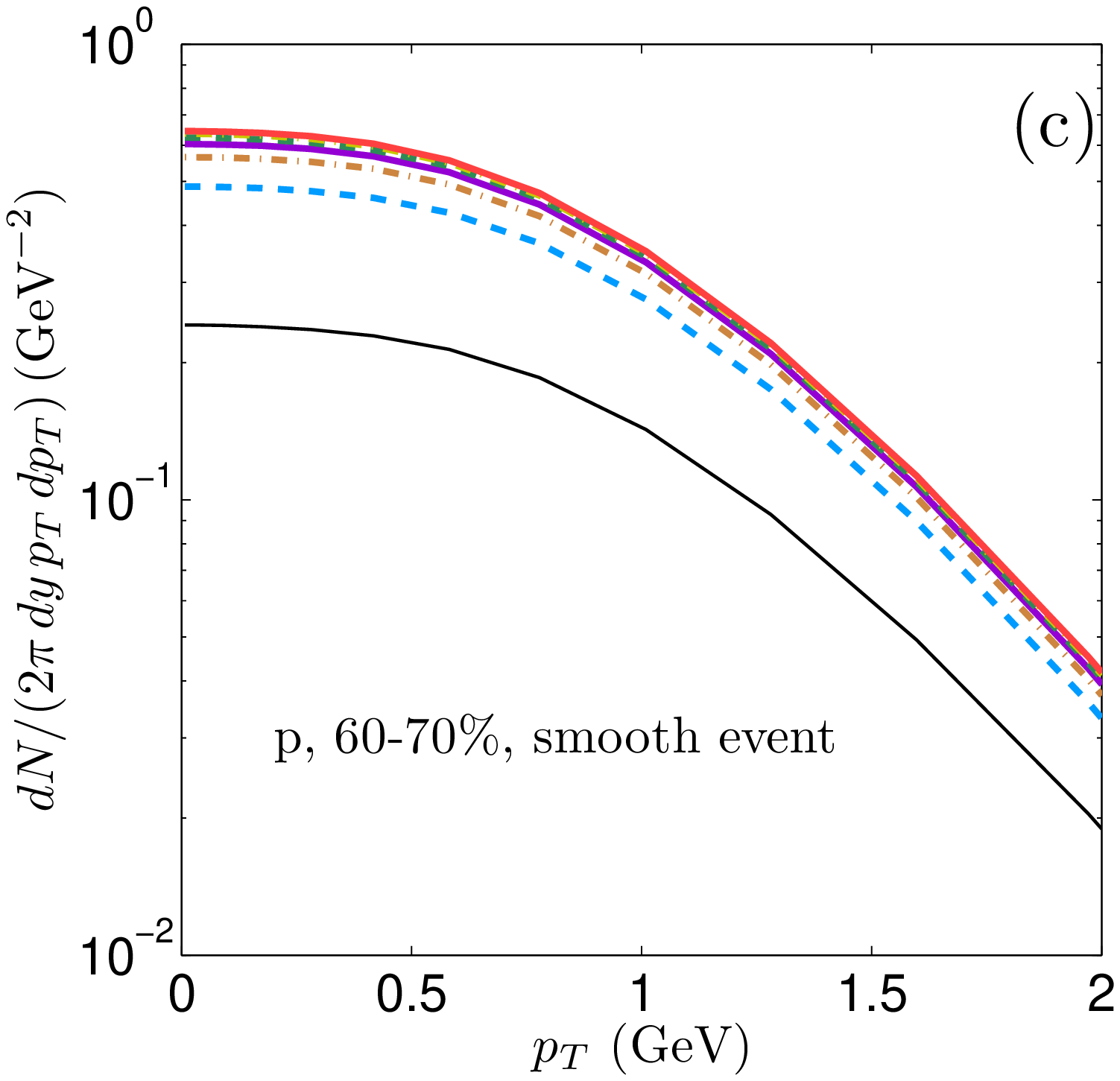}
	  \includegraphics[width=8cm,height=7cm]{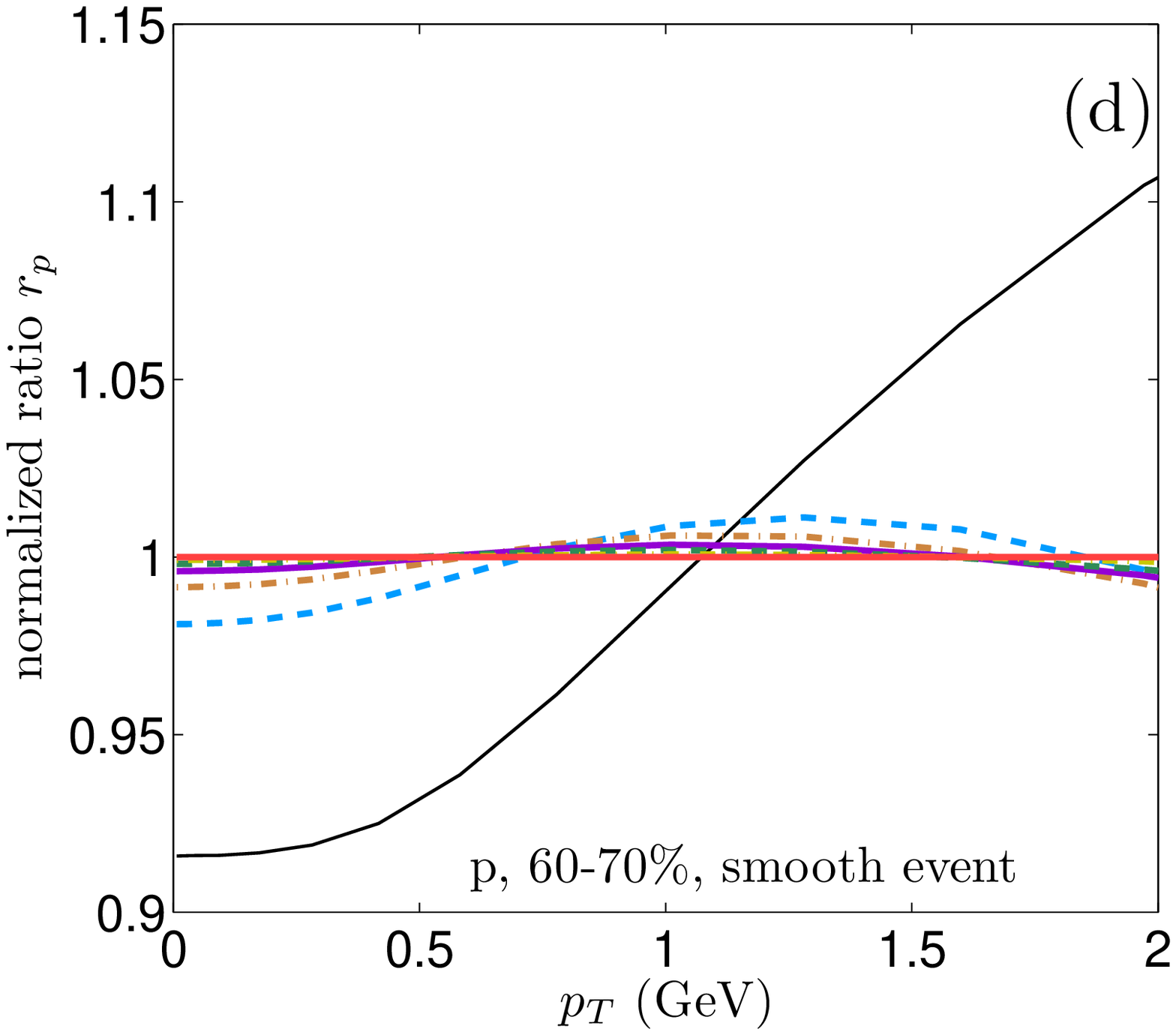}
  }
  \caption{(Color online) Same as Fig.\ref{fig:1}, but for protons. 
  \label{fig:2}}
\end{figure*}
%

\section{Results and discussion}
\label{sec:3}

Using the ordered tables described in Sec.~\ref{sec:2} and truncating the sum over resonance decay contributions at $j_\mathrm{cut}$ values corresponding to various different cumulative resonance decay contribution thresholds $c_i^\mathrm{cut}$, we performed calculations for $\pi^+$, $K^+$, and $p$. We tested individual bumpy as well as (ellipticity-aligned and ensemble-averaged) smooth initial conditions at both RHIC and LHC energies, for a variety of collision centralities. Since the results were found all to be qualitatively similar, we show only a small selection, focussing on pions and protons from one bumpy Pb-Pb event from the $0{-}10\%$ centrality class and from the smooth averaged initial condition corresponding to the $60{-}70\%$ centrality class, both at LHC energy ($\sqrt{s}{\,=\,}2.76\,A$\,GeV).

Figure \ref{fig:1} shows the pion $p_T$-spectra, for the bumpy central collision in the upper panels and the smooth peripheral event in the lower panels. The left panels show the usual semilogarithmic plots of the absolutely normalized $p_T$-distribution. As is well-known, the directly emitted (``thermal'') pions constitute only about 50-60\% of all observed pions, the rest coming from resonance decays. The ``thermal" spectrum also has the wrong shape: resonance decay pions predominantly contribute to the low-$p_T$ part of the spectrum, making it steeper. However, this shape difference between the truncated and full resonance decay spectrum disappears almost completely already when including only the 9 strongest decay channels, accounting for just 60\% of the total pion yield from resonance decays. This is shown in the right panels of Fig.~\ref{fig:1} where we plot the ratio 
\begin{equation}
\label{ratio}
r_i(p_T) = \frac{
                 \frac{dN_i^\mathrm{th}}{dyp_Tdp_T} + 
                 \frac{1}{\sum_{j{=}1}^{j_\mathrm{cut}}c_{j{\to}i}} \sum_{j{=}1}^{j_\mathrm{cut}}
                 \frac{dN_i^{(j)}}{dyp_Tdp_T}}
                 {\frac{dN_i^\mathrm{th}}{dyp_Tdp_T} + 
                 \sum_{j{=}1}^{j_\mathrm{max}}
                 \frac{dN_i^{(j)}}{dyp_Tdp_T}}
\end{equation}
for $i{\,=\,}\pi$ as a function of $p_T$. ($N_i^{(j)}$ is the contribution to particle species $i$ from decays of particle species $j$ (see Eq.~(\ref{eq2})), and $j_\mathrm{max}$ is the index of the last resonance in 
%
\begin{figure*}[ht]
  \center{
	  \includegraphics[width=8cm, height=5.5cm]{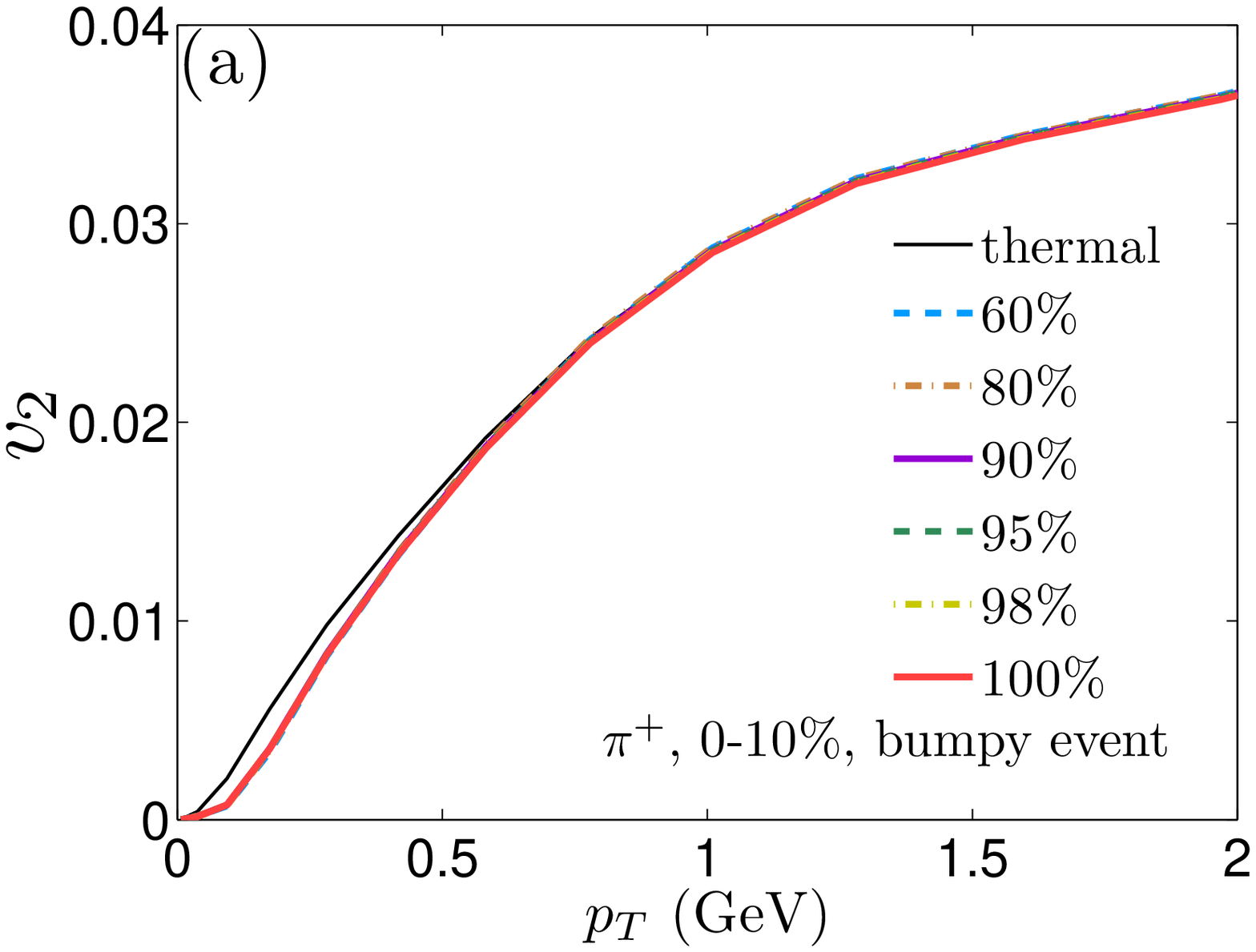}
	  \includegraphics[width=8cm, height=5.5cm]{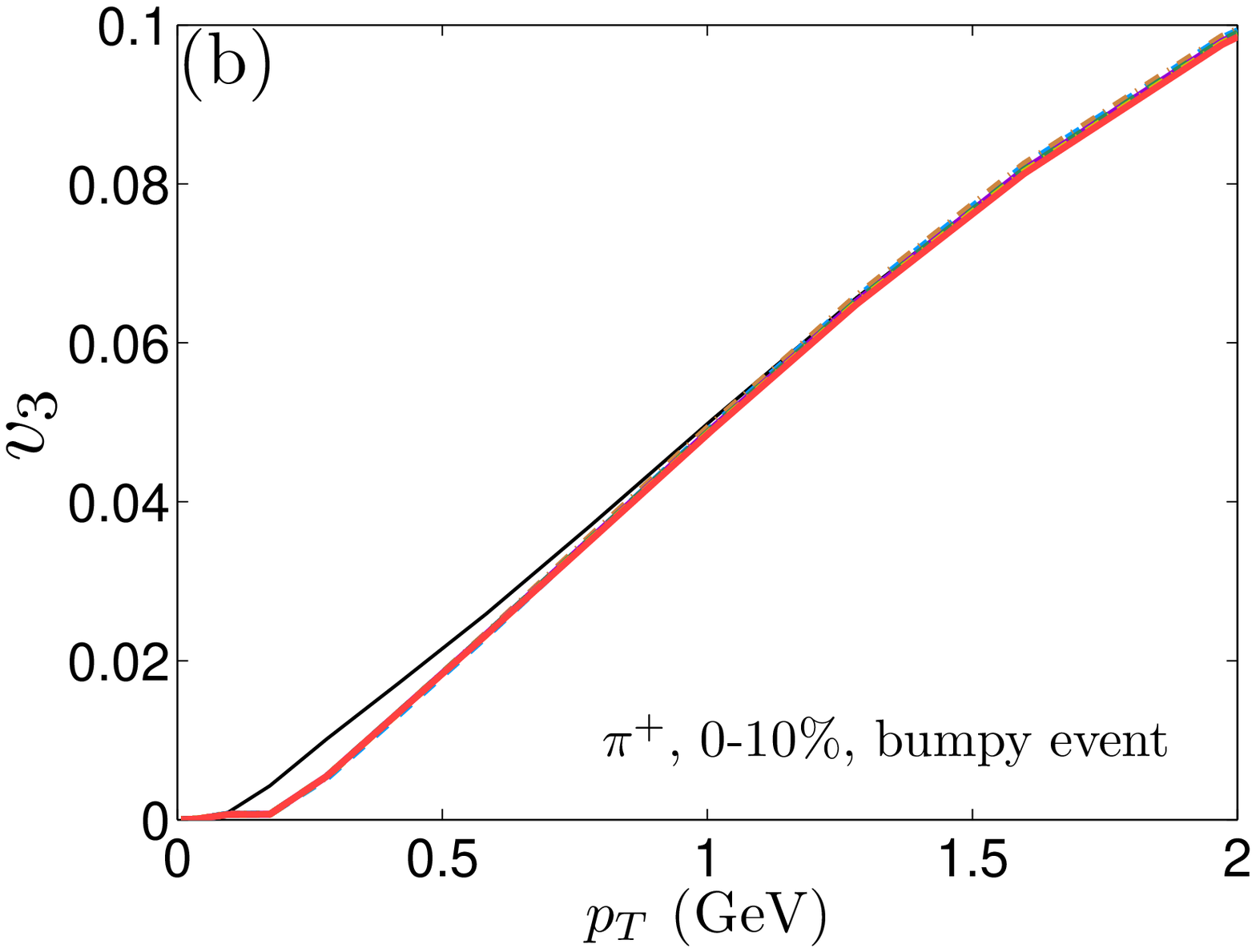}\\
	  \includegraphics[width=8cm, height=5.5cm]{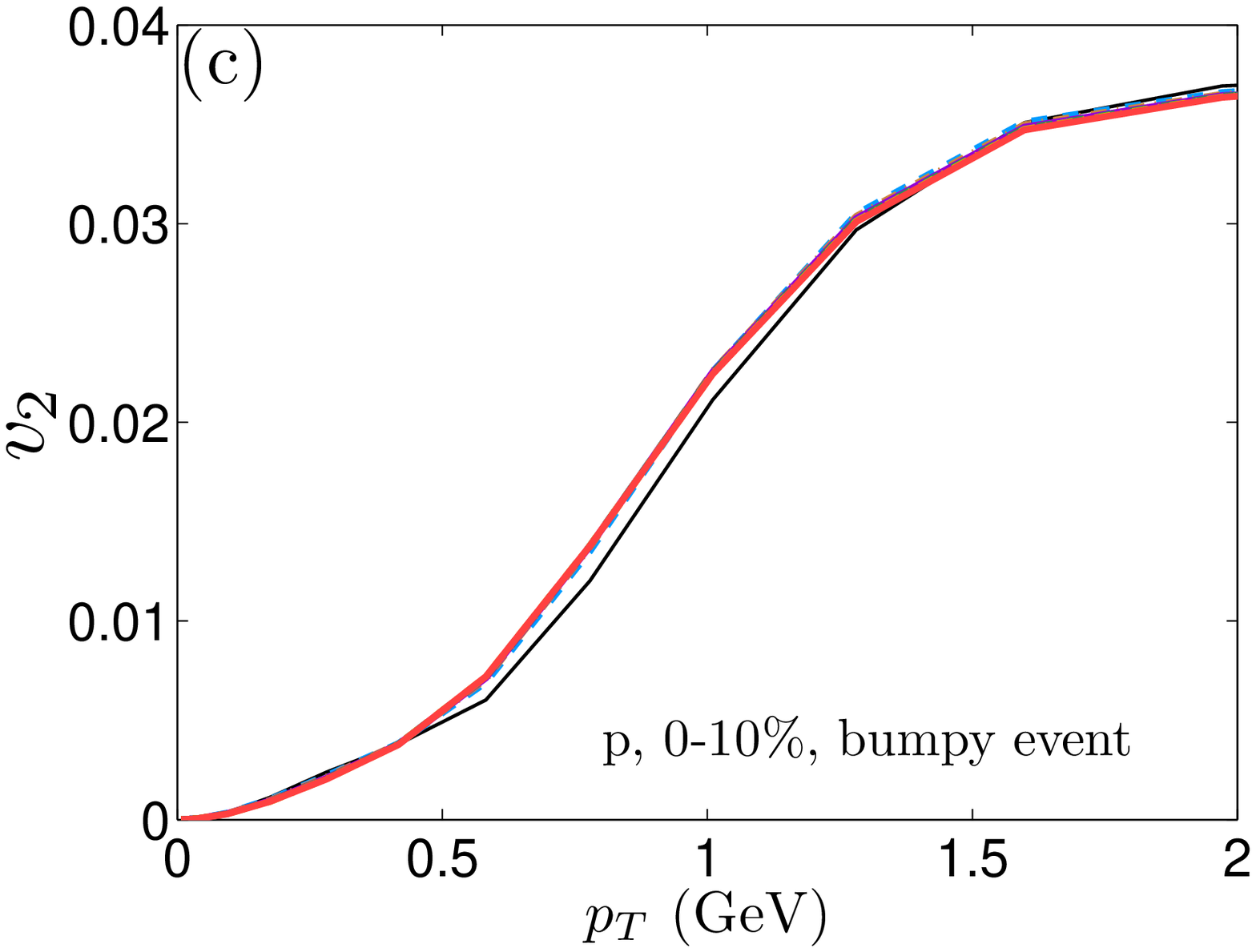}
	  \includegraphics[width=8cm, height=5.5cm]{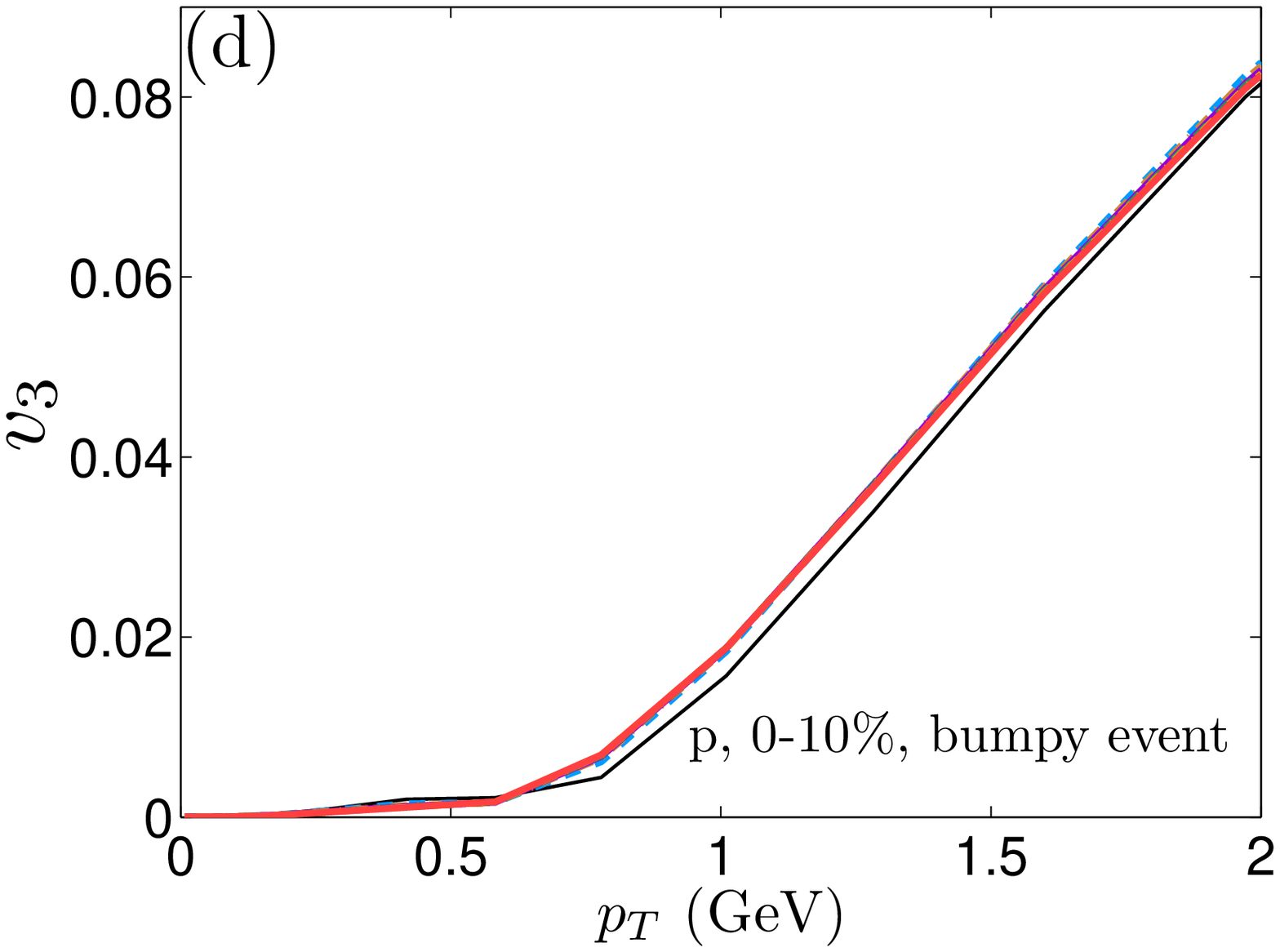}
  }
  \caption{(Color online) The differential elliptic ($v_2$, left panel) and triangular flow ($v_3$,
   right panel), for $\pi^+$ (upper panels) and $p$ (lower panels), for one bumpy Pb-Pb 
   event from the $0{-}10\%$ centrality class at LHC energy. As in Fig.~\ref{fig:1}, lines of
   different styles and colors correspond to different cumulative resonance decay fractions.
  \label{fig:3}}
\end{figure*}
%
the ordered resonance decay table from Sec.~\ref{sec:2}.) The numerator includes only resonance decays up to $j_\mathrm{cut}$, but we renormalize those decay contributions by the cumulative decay contribution $c_i^\mathrm{cut}$ corresponding to the same $j_\mathrm{cut}$ value. ($c_i^\mathrm{cut}$ is easily calculated from Eqs.~(\ref{eq2},\ref{eq3}) and directly obtained by summing the entries in the third column of the resonance decay table.) This renormalization corrects for the missing yield from the truncation of the decay table. The remaining effect (after missing yield renormalization) of the truncation on the {\em shape} of the $p_T$-spectrum is seen in panels (b) and (d) of Fig.~\ref{fig:1}: Whereas without any resonance decays the ratio $r_\pi(p_T)$ changes by almost a factor 2 between $p_T{\,=\,}0$ and 2\,GeV, this variation is reduced to less than 5\% already for $c_\pi^\mathrm{cut}{\,=\,}60\%$, for both bumpy and smooth initial conditions in both central and peripheral collisions. 

In Fig.~\ref{fig:2} we show in the same way the proton spectra. Again the shape of the spectra can be accurately reproduced by taking into account a small fraction of all decay contributions (note the expanded vertical scale in Figs.~\ref{fig:2}b,d): after renormalization to account for the missing yield, just the 4 strongest of 75 decay channels (three charge states of the $\Delta(1232)$ resonance and one charge state of $\Delta(1600)$), corresponding to 60\% of the total resonance decay yield for protons, reproduce the full proton spectrum with ${<\,}5\%$ error between $p_T{\,=\,}0$ and 2\,GeV.

We conclude that, by accounting for the missing yield through appropriate renormalization, the correctly normalized total pion and proton spectra can be obtained, with shape errors ${<\,}5\%$, by including only the strongest decay channels accounting for the leading 60\% of the total resonance decay yields. A quick look at the tables in the Appendix shows that this will reduce the number of resonance decays (and thus computer time) by at least a factor 10.

We now proceed to a discussion of the differential and $p_T$-integrated anisotropic flow coefficients $v_n$, defined by
\begin{eqnarray}
\label{eq5}
  &&v_n(p_T)\, e^{in\psi_n(p_T)} =
  \left. \!\!\int\!\! d\phi_p\, e^{in\phi_p}\, \frac{dN}{dy p_T dp_T d\phi_p} \right /
  \frac{dN}{dy p_T dp_T}\,,
\nonumber\\
  &&v_n\, e^{in\psi_n} =
  \left. \int p_T\, dp_T\,d\phi_p\, e^{in\phi_p}\, \frac{dN}{dy p_T dp_T d\phi_p} \right /
  \frac{dN}{dy}\,.
\end{eqnarray}
Here the spectrum $dN/(dy p_T dp_T d\phi_p)$ includes all contributions from the ordered resonance decay table for the considered stable species up to a 
%
\begin{figure*}[ht]
  \center{
	  \includegraphics[width=8cm, height=5.6cm]{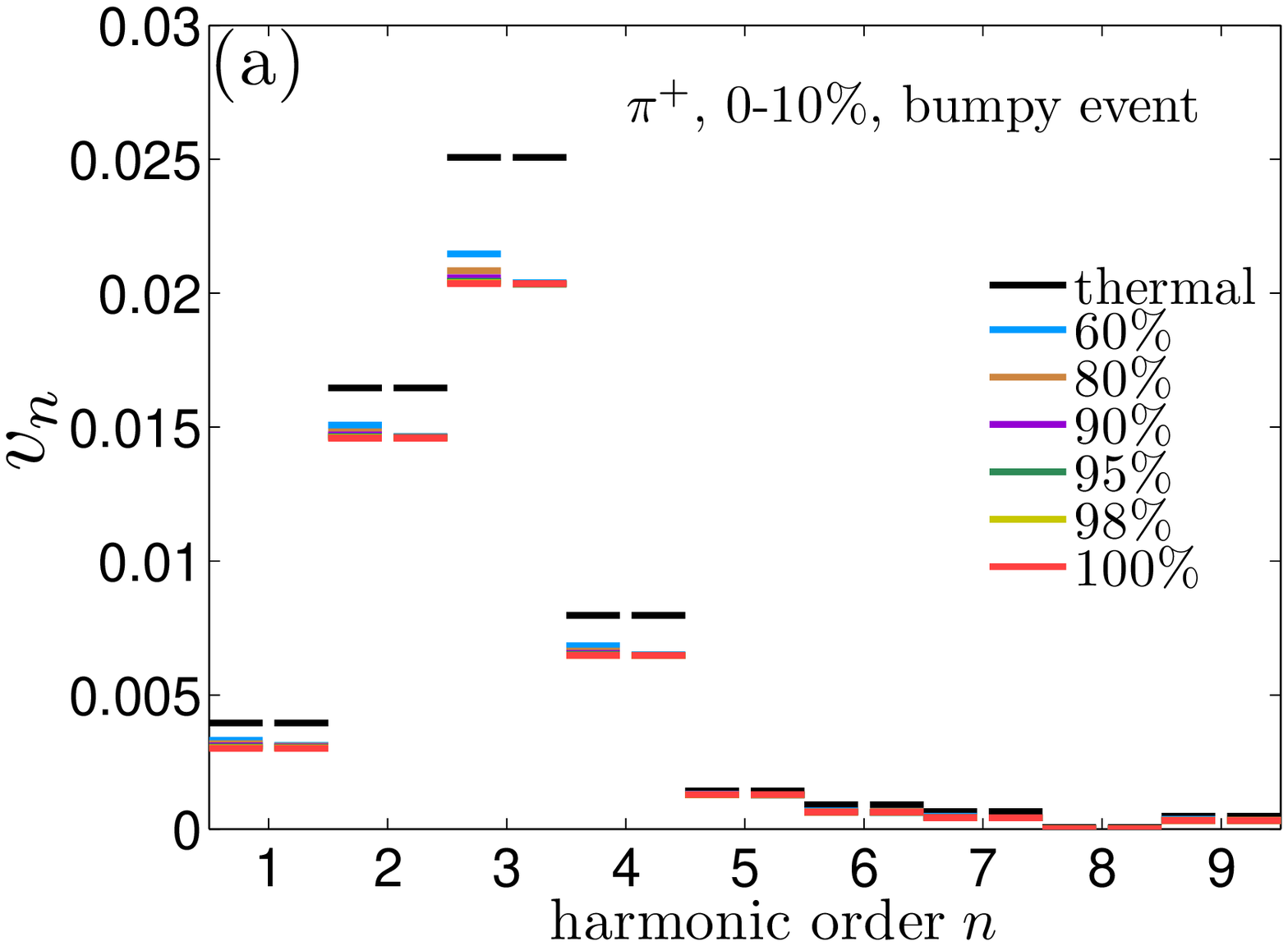}
	  \includegraphics[width=8cm, height=5.6cm]{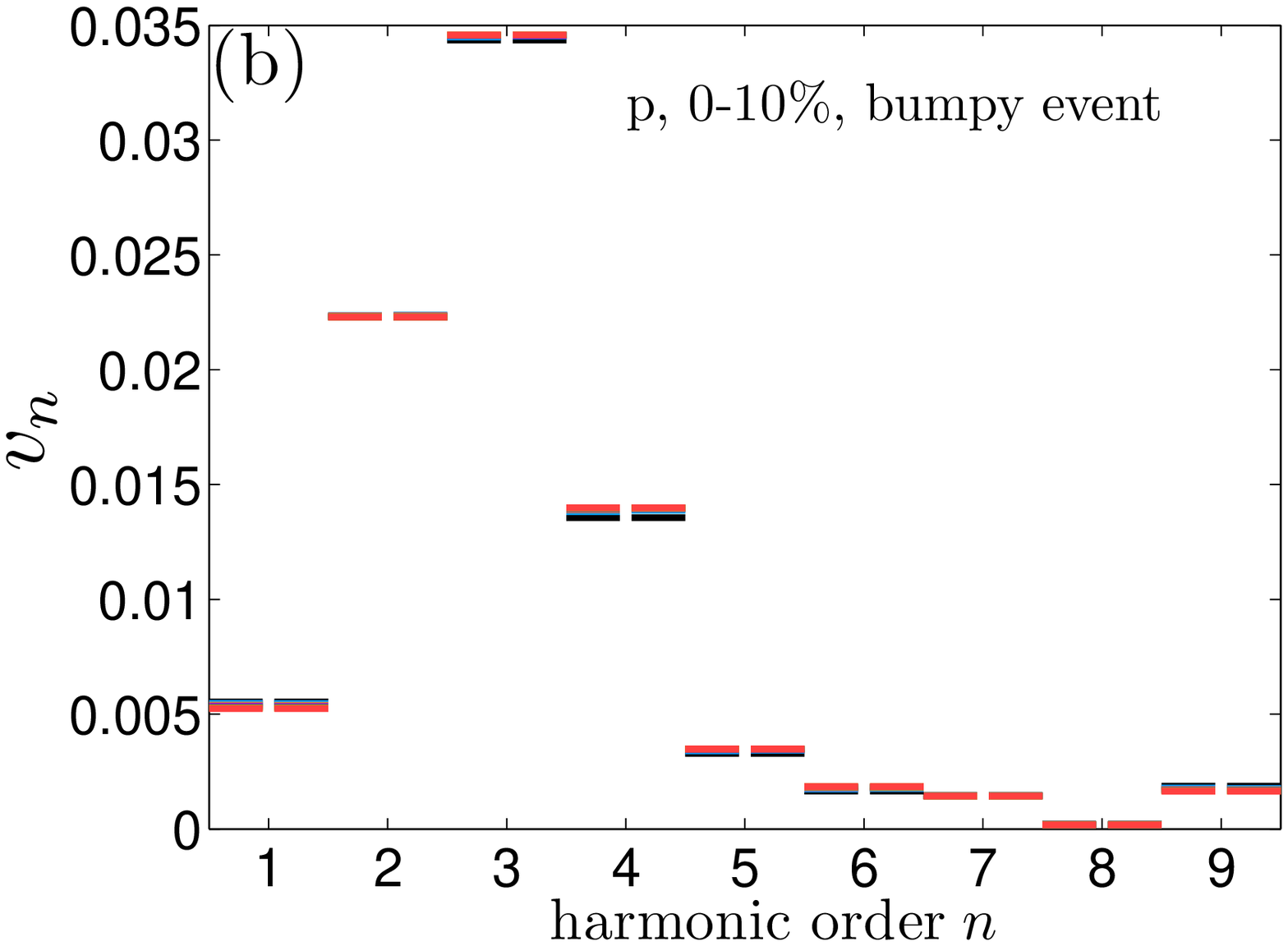}\\
	  \includegraphics[width=8cm, height=5.6cm]{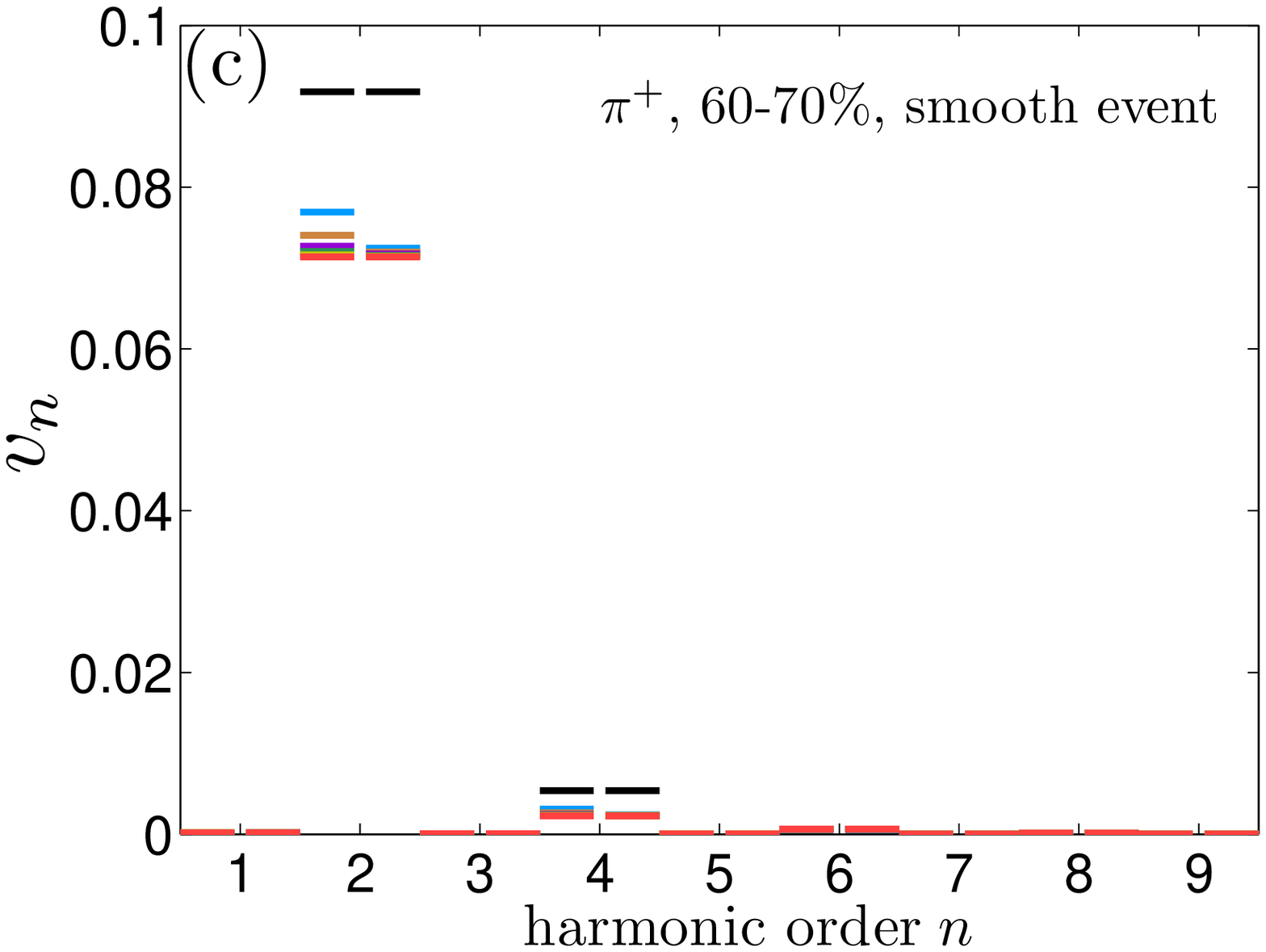}
	  \includegraphics[width=8cm, height=5.6cm]{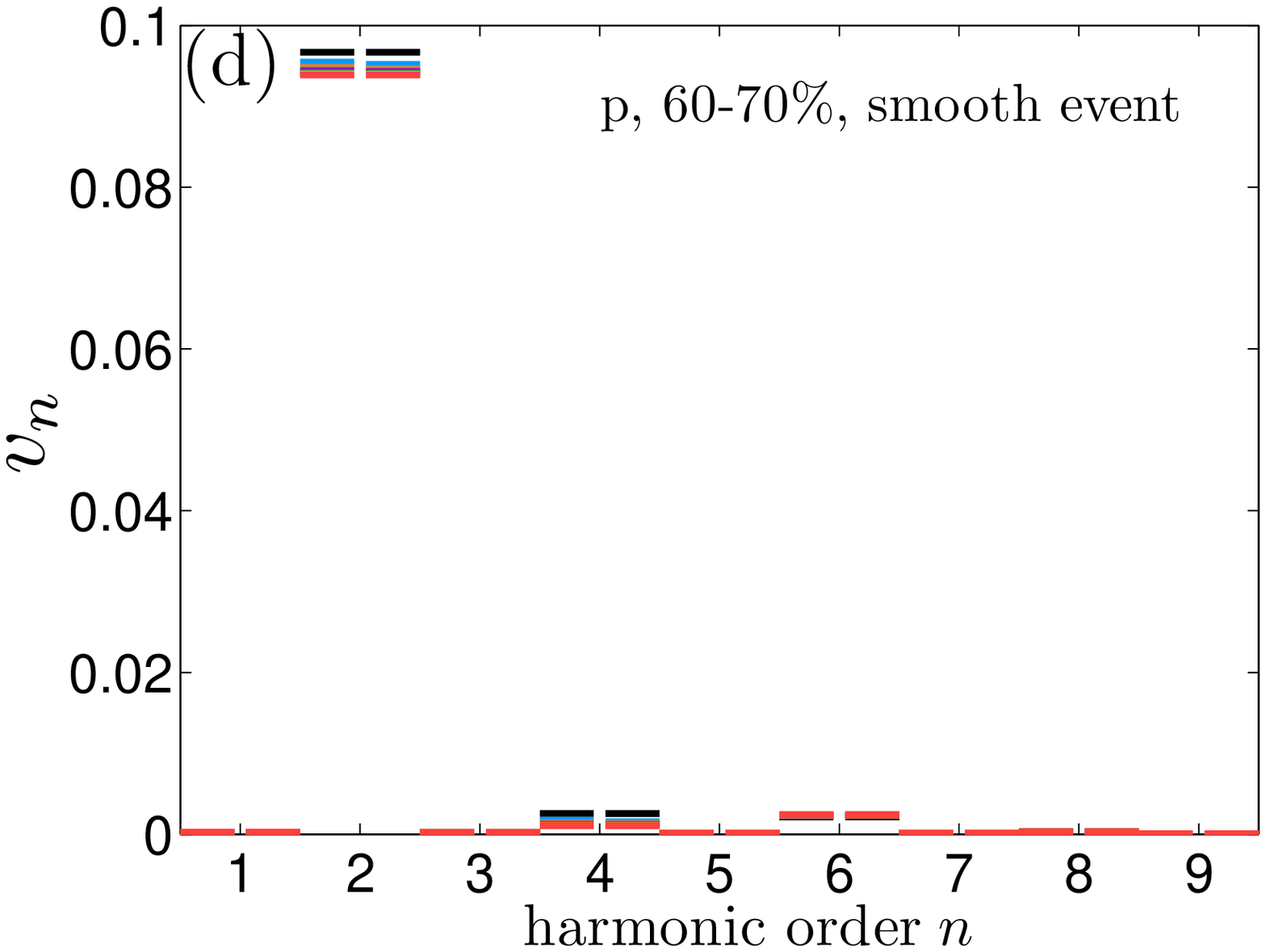}
  }
  \caption{(Color online) The $p_T$-integrated anisotropic flow coefficients $v_n$, 
  $n{\,=\,}1,\dots,9$, for $\pi^+$ (a,c) and $p$ (b,d), for a bumpy central (0-10\%) collision 
  event (a,b) and a smooth averaged peripheral (60-70\%) collision event (c,d). Line colors 
  and styles as in Figs.~\ref{fig:1}-\ref{fig:3}. For a discussion of the two sets of lines for 
  each harmonic order $n$ see text.
  \label{fig:4}}
\end{figure*}
%
certain threshold $j_\mathrm{cut}$, with the truncated resonance decay contribution renormalized for the missing yield by a factor $1/c_i^\mathrm{cut}$ as shown in the numerator of Eq.~(\ref{ratio}). In Figs.~\ref{fig:3} and \ref{fig:4} below we specify the cumulative decay contribution percentage $c_i^\mathrm{cut}$ to indicate the truncation level corresponding to each curve.

Figure~\ref{fig:3} shows the differential elliptic and triangular flows for pions and protons, for one single bumpy central (0-10\% centrality) event. We see that once again excellent agreement with the full resonance decay calculation is already obtained when including only the small subset of resonances that account for the top 60\% of the resonance decay yields. We checked that this result is generic, i.e. it does not depend on the selected event (although the elliptic and triangular flows do).

For the $p_T$-integrated harmonic flow coefficients $v_n$ we show in Fig.~\ref{fig:4} results for all harmonic orders from $n{\,=\,}1$ to 9, again for pions and protons and for a bumpy central as well as a smooth peripheral event. For the smooth averaged initial condition, the odd harmonics vanish by symmetry. For fluctuating initial conditions, the $v_n$ values shown here and their relative size depend on the randomly selected event. All plots shown in this paper are based on one and the same bumpy central collision event. 

For each harmonic order $n$, Fig.~\ref{fig:4} shows two sets of results. The left set corresponds to results obtained by using the truncated resonance decay spectra shown in Figs.~\ref{fig:1}a,c and \ref{fig:2}a,c, without missing yield renormalization. The right set
uses the renormalized truncated decay spectra as defined in the numerator of Eq.~(\ref{ratio}). One observes a much faster convergence towards the full result in the right sets than in the left sets. The reason is that, by renormalizing the truncated resonance decay contributions for the missing yield, the correct mixing ratio between direct thermal and indirect decay contributions is ensured and the shape of the total $p_T$-spectrum is approximated much more accurately than without renormalization (see Figs.~\ref{fig:1}b,d and \ref{fig:2}b,d). Figure~\ref{fig:4} demonstrates that, when using the renormalized truncated decay spectra, accounting for just the top 60\% decay contributions (i.e. including only the 9 strongest decay channels contributing to pions and the 4 strongest decay channels contributing to the proton spectra) reproduces the full results for the harmonic flow coefficients $v_n$ with excellent precision: The lines corresponding to different $c_i^\mathrm{cut}$ values ${\geq}60\%$ are almost indistinguishable.   

Future precision extractions of the QGP viscosity may require highly precise $v_n$ values. For such a purpose one can adjust $j_\mathrm{cut}$ to include a larger fraction of all resonance decays if needed.

For a given precision, the required minimal $j_\mathrm{cut}$ truncation indices and cumulative resonance decay fractions $c_\mathrm{K}^\mathrm{cut}$ for kaons lie between those for pions and protons. The $c_i^\mathrm{cut}$ for $i{\,=\,}\pi$,\,K,\,$p$ are almost identical at RHIC and LHC energies, i.e. only weakly sensitive to radial flow.

\section{Conclusions}
\label{sec:4}

We have shown that for a sufficiently accurate determination of the differential anisotropic flow coefficients $v_n(p_T)$ only those resonances need to be included that generate the top 60\% of the largest decay contributions to the stable particle yields. For the single particle spectra, correct normalization of the total yield requires a renormalization of the truncated resonance decay yield as given in the numerator of Eq.~(\ref{ratio}). With this renormalization, good convergence of the slope of the pion spectra and of the $p_T$-integrated anisotropic flow coefficients $v_n$ requires inclusion of only the 9 strongest contributing channels for pions and only the 4 strongest channels for protons, accounting in both cases for just 60\% of the total decay yield. This reduces the number of resonance decay channels to be evaluated by a factor ${>}10$, without loss of precision, leading to a similar reduction of the total computing time for the final stable hadron distributions.   

In hybrid model calculations \cite{Song:2010aq} the late hadronic stage is described microscopically by a Boltzmann cascade that propagates a reduced set of resonances until final kinetic decoupling. In this case the spectra of all unstable resonances that are explicitly included in the Boltzmann cascade must be generated on the conversion surface. This is still only a small subset of all resonances included in the resonance decay tables. The optimal ordering of the resonance decay tables for the purpose of generating input for the late-stage Boltzmann cascade and the corresponding optimized truncation fractions $c_i^\mathrm{cut}$ will be studied in a follow-up report. 

{\sl Acknowledgments:}
This work was supported by the U.S.\ Department of Energy under Grants No.~\rm{DE-SC0004286} and (within the framework of the JET Collaboration) \rm{DE-SC0004104}.


\section{Appendix: Feed down contribution tables for $\pi$, $K$, $p$, $\Lambda$, $\Sigma^+$, and $\Xi^-$}
\label{sec:6}
%
\begin{table}[h]
\begin{tabular}{c|c|c|c}
\hline \hline
name & mass (GeV) & total contribution (\%)  &  \\ \hline
$\omega$  &     0.78259 &     15.398  &    \\
$\rho^{0}$  &      0.7758 &     11.179  &    \\
$\rho^{+}$  &      0.7758 &     11.098  &    \\
$K^{*+}(892)$  &     0.89166 &       5.54  &    \\
$\bar{K}^{*0}(892)$  &      0.8961 &      5.355  &    \\
$\eta$  &     0.54775 &      4.682  &    \\
$\bar{\Delta}^{-}  (1232)$  &       1.232 &      2.613  &    \\
$\Delta^{++} (1232) $  &       1.232 &      2.606  &    \\
$b_1^{+} (1235) $  &      1.2295 &      2.498  &  60\%  \\ \hline
$\eta'(958)$  &     0.95778 &      2.069  &    \\
$a_0^{+}(980)$  &      0.9847 &      1.862  &    \\
$h_1 (1170)$  &        1.17 &       1.26  &    \\
$a_1^{+} (1260) $  &        1.23 &      1.226  &    \\
$b_1^{-} (1235) $  &      1.2295 &       1.19  &    \\
$b_1^{0} (1235) $  &      1.2295 &      1.181  &    \\
$a_2^{+} (1320) $  &      1.3183 &      1.177  &    \\
$\Sigma^{+} (1385) $  &      1.3828 &       1.09  &    \\
$\bar{\Sigma}^{-} (1385) $  &      1.3872 &      1.057  &    \\
$f_1 (1285)$  &      1.2818 &      0.994  &    \\
$\bar{K}_1^{0} (1270) $  &       1.273 &      0.963  &    \\
$K_1^{+} (1270) $  &       1.273 &      0.962  &    \\
$\bar{\Delta}^{0} (1232) $  &       1.232 &      0.857  &    \\
$\Delta^{+} (1232) $  &       1.232 &      0.856  &    \\
$a_1^{0} (1260) $  &        1.23 &      0.817  &    \\
$a_2^{0} (1320) $  &      1.3183 &      0.771  &    \\
$\phi(1020)$  &      1.0195 &      0.762  &  80\%  \\ \hline
$f_0(980)$  &      0.9741 &      0.613  &    \\
$K_1^{0} (1270) $  &       1.273 &      0.472  &    \\
$K_1^{-} (1270) $  &       1.273 &      0.472  &    \\
$f_2 (1270)$  &      1.2754 &       0.45  &    \\
$a_1^{-} (1260) $  &        1.23 &      0.409  &    \\
$a_0^{-}(980)$  &      0.9847 &      0.402  &    \\
$a_0^{0}(980)$  &      0.9847 &      0.399  &    \\
$a_2^{-} (1320) $  &      1.3183 &      0.398  &    \\
$\pi_1^{+} (1400) $  &       1.376 &      0.373  &    \\
$K_2^{*+} (1430) $  &      1.4256 &      0.368  &    \\
$\Xi^{0} (1530)$  &      1.5318 &      0.357  &    \\
$\bar{\Delta}^{-} (1600) $  &         1.6 &      0.356  &    \\
$\Delta^{++} (1600) $  &         1.6 &      0.356  &    \\
$\bar{K}_2^{*0} (1430) $  &      1.4324 &       0.35  &    \\
$\bar{\Xi}^{-} (1530)$  &       1.535 &      0.347  &    \\
$\bar{K}_1^{0} (1400) $  &       1.402 &      0.341  &    \\
$K_1^{+} (1400) $  &       1.402 &       0.34  &    \\
$p (1520)$  &        1.52 &      0.305  &    \\
$\bar{n} (1520)$  &        1.52 &      0.305  &    \\
$\eta (1295)$  &       1.294 &      0.297  &    \\
$\bar{K}^{*-} (1410) $  &       1.414 &      0.294  &    \\
$K^{*+} (1410) $  &       1.414 &      0.294  &    \\
\hline \hline
\end{tabular}
\caption{Resonance contribution list for $\pi^+$, for $T_\mathrm{conv}{\,=\,}120$\,MeV.}
\label{table1_1}
\end{table}

\begin{table}[h]
\begin{tabular}{c|c|c|c}
\hline \hline
name & mass (GeV) & total contribution (\%) &  \\ \hline
$\pi^{+} (1300) $  &         1.3 &      0.257  &    \\
$\bar{\Delta}^{0} (1600) $  &         1.6 &      0.249  &    \\
$\Delta^{+} (1600) $  &         1.6 &      0.249  &    \\
$\bar{n} (1440)$  &        1.44 &      0.241  &    \\
$p (1440)$  &        1.44 &      0.241  &    \\
$\omega (1420)$  &       1.419 &      0.205  &    \\
$a_0^{+} (1450) $  &       1.474 &      0.196  &    \\
$\Lambda (1405)$  &      1.4065 &      0.188  &  90\%  \\ \hline
$\bar{\Lambda} (1405)$  &      1.4065 &      0.188  &    \\
$\eta_2 (1645)$  &       1.617 &      0.185  &    \\
$f_1 (1420)$  &      1.4263 &      0.174  &    \\
$K_1^{0} (1400) $  &       1.402 &      0.173  &    \\
$K_1^{-} (1400) $  &       1.402 &      0.173  &    \\
$a_0^{0} (1450) $  &       1.474 &      0.153  &    \\
$\bar{n} (1675)$  &       1.675 &      0.147  &    \\
$p (1675)$  &       1.675 &      0.147  &    \\
$\bar{\Delta}^{-} (1700) $  &         1.7 &      0.145  &    \\
$\Delta^{++} (1700) $  &         1.7 &      0.145  &    \\
$\omega_3 (1670)$  &       1.667 &      0.144  &    \\
$\pi^{0} (1300) $  &         1.3 &      0.142  &    \\
$K^{*0} (1410) $  &       1.414 &       0.14  &    \\
$K^{*-} (1410) $  &       1.414 &       0.14  &    \\
$\Delta^{0} (1600) $  &         1.6 &      0.135  &    \\
$\bar{\Delta}^{+} (1600) $  &         1.6 &      0.134  &    \\
$\rho_3^{+} (1690) $  &      1.6888 &      0.123  &    \\
$p (1680)$  &       1.685 &      0.119  &    \\
$\bar{n} (1680)$  &       1.685 &      0.119  &    \\
$\bar{\Delta}^{0} (1700) $  &         1.7 &      0.117  &    \\
$\Delta^{+} (1700) $  &         1.7 &      0.117  &    \\
$K_2^{*-} (1430) $  &      1.4256 &      0.115  &    \\
$\bar{\Delta}^{-} (1620) $  &        1.63 &      0.113  &    \\
$\Delta^{++} (1620) $  &        1.63 &      0.113  &    \\
$\Lambda (1520)$  &      1.5195 &      0.112  &    \\
$\bar{\Lambda} (1520)$  &      1.5195 &      0.112  &    \\
$K_2^{*0} (1430) $  &      1.4324 &      0.109  &    \\
$\eta (1405)$  &      1.4103 &      0.104  &    \\
$\bar{n} (1535)$  &       1.535 &      0.103  &    \\
$p (1535)$  &       1.535 &      0.102  &    \\
$\bar{n} (1700)$  &         1.7 &      0.102  &    \\
$p (1700)$  &         1.7 &      0.102  &    \\
$\bar{n} (1720)$  &        1.72 &      0.098  &    \\
$p (1720)$  &        1.72 &      0.098  &    \\
$\rho_3^{0} (1690) $  &      1.6888 &      0.097  &    \\
$\pi^{-} (1300) $  &         1.3 &      0.093  &    \\
$a_0^{-} (1450) $  &       1.474 &      0.091  &    \\
$\bar{p} (1520)$  &        1.52 &      0.088  &   95\% \\ \hline
$n (1520)$  &        1.52 &      0.088  &    \\
$\bar{\Delta}^{0} (1620)$  &        1.63 &      0.086  &    \\
$\Delta^{+} (1620) $  &        1.63 &      0.086  &    \\
$\pi_2^{+} (1670)  $  &      1.6724 &      0.086  &    \\
\hline \hline
\end{tabular}
\caption{Resonance contribution list for $\pi^+$, for $T_\mathrm{conv}{\,=\,}120$\,MeV (continued).}
\label{table1_2}
\end{table}

\begin{table}[h]
\begin{tabular}{c|c|c|c}
\hline \hline
name & mass (GeV) & total contribution (\%) &  \\ \hline
$\rho^{0}  (1450) $  &       1.465 &      0.082  &    \\
$\rho^{+}  (1450) $  &       1.465 &      0.082  &    \\
$\pi_1^{-} (1400) $  &       1.376 &       0.08  &    \\
$\pi_1^{0} (1400) $  &       1.376 &       0.08  &    \\
$\bar{\Sigma}^{-} (1670)$  &        1.67 &      0.079  &    \\
$\Sigma^{+} (1670)$  &        1.67 &      0.079  &    \\
$\Sigma^{+} (1775)$  &       1.775 &      0.075  &    \\
$\bar{\Sigma}^{-} (1775)$  &       1.775 &      0.075  &    \\
$\rho^{+} (1700)$  &        1.72 &      0.072  &    \\
$n (1700)$  &         1.7 &      0.072  &    \\
$\bar{p} (1700)$  &         1.7 &      0.072  &    \\
$n (1440)$  &        1.44 &      0.072  &    \\
$\bar{p} (1440)$  &        1.44 &      0.072  &    \\
$\Delta^{0} (1700) $  &         1.7 &       0.07  &    \\
$\bar{\Delta}^{+} (1700) $  &         1.7 &       0.07  &    \\
$\Lambda (1690)$  &        1.69 &      0.069  &    \\
$\bar{\Lambda} (1690)$  &        1.69 &      0.069  &    \\
$\Sigma^{0} (1385) $  &      1.3837 &      0.069  &    \\
$\bar{\Sigma}^{0} (1385) $  &      1.3837 &      0.069  &    \\
$\omega (1650)$  &        1.67 &      0.067  &    \\
$n (1675)$  &       1.675 &      0.061  &    \\
$\bar{p} (1675)$  &       1.675 &      0.061  &    \\
$\pi_1^{+} (1600) $  &       1.653 &       0.06  &    \\
$\bar{K}_2^{0} (1770) $  &       1.773 &      0.056  &    \\
$K_2^{+} (1770) $  &       1.773 &      0.056  &    \\
$\Sigma^{0} (1670)$  &        1.67 &       0.05  &    \\
$\bar{\Sigma}^{0} (1670)$  &        1.67 &       0.05  &    \\
$\bar{\Delta}^{-} (1905) $  &        1.89 &       0.05  &    \\
$\Delta^{++} (1905) $  &        1.89 &       0.05  &    \\
$\rho^{0} (1700) $  &        1.72 &      0.049  &    \\
$\Delta^{0} (1620) $  &        1.63 &      0.048  &    \\
$\bar{\Delta}^{+} (1620) $  &        1.63 &      0.048  &    \\
$\pi_2^{0} (1670)  $  &      1.6724 &      0.045  &    \\
$\bar{n} (1710)$  &        1.71 &      0.044  &    \\
$\bar{\Delta}^{0} (1905) $  &        1.89 &      0.044  &    \\
$K_0^{*+} (1430) $  &       1.412 &      0.044  &    \\
$\Delta^{+} (1905) $  &        1.89 &      0.044  &    \\
$p (1710)$  &        1.71 &      0.044  &    \\
$\bar{K}_0^{*0} (1430) $  &       1.412 &      0.044  &    \\
$\bar{\Sigma}^{-} (1660)$  &        1.66 &      0.043  &    \\
$\Sigma^{+} (1660)$  &        1.66 &      0.043  &    \\
$\bar{n} (1650)$  &       1.655 &      0.042  &    \\
$p (1650)$  &       1.655 &      0.042  &    \\
$f_0 (1500)$  &       1.507 &      0.041  &    \\
$\eta (1475)$  &       1.476 &       0.04  &  98\%  \\ \hline
$\bar{\Delta}^{-}(1950) $  &        1.93 &       0.04  &    \\
$\Delta^{++} (1950) $  &        1.93 &       0.04  &    \\
$\bar{\Delta}^{-} (1920) $  &        1.92 &      0.039  &    \\
\hline \hline
\end{tabular}
\caption{Resonance contribution list for $\pi^+$, for $T_\mathrm{conv}{\,=\,}120$\,MeV (continued).}
\label{table1_3}
\end{table}

\begin{table}
\begin{tabular}{c|c|c|c}
\hline \hline
name & mass (GeV) & total contribution (\%) &  \\ \hline
$\Delta^{++}(1920) $  &        1.92 &      0.039  &    \\
$\Lambda (1830)$  &        1.83 &      0.037  &    \\
$\bar{\Lambda} (1830)$  &        1.83 &      0.037  &    \\
$\rho_3^{-} (1690) $  &      1.6888 &      0.037  &    \\
$\bar{K}_3^{*0} (1780) $  &       1.776 &      0.034  &    \\
$K_3^{*+} (1780)$  &       1.776 &      0.034  &    \\
$\rho^{-} (1700) $  &        1.72 &      0.034  &    \\
$n (1535)$  &       1.535 &      0.033  &    \\
$\bar{p} (1535)$  &       1.535 &      0.033  &    \\
$n (1720)$  &        1.72 &      0.033  &    \\
$\bar{p} (1720)$  &        1.72 &      0.033  &    \\
$\bar{p} (1680)$  &       1.685 &      0.032  &    \\
$n (1680)$  &       1.685 &      0.032  &    \\
$K_2^{0} (1770) $  &       1.773 &      0.032  &    \\
$\bar{K}_2^{0} (1820) $  &       1.816 &      0.032  &    \\
$K_2^{-} (1770) $  &       1.773 &      0.032  &    \\
$K_2^{+} (1820) $  &       1.816 &      0.032  &    \\
$\pi_2^{-} (1670)  $  &      1.6724 &      0.031  &    \\
$\Lambda (1600)$  &         1.6 &       0.03  &    \\
$\bar{\Lambda} (1600)$  &         1.6 &       0.03  &    \\
$f_0 (1370)$  &         1.4 &      0.029  &    \\
$\Delta^{0} (1905) $  &        1.89 &      0.028  &    \\
$\bar{\Delta}^{+} (1905) $  &        1.89 &      0.028  &    \\
$n (1710)$  &        1.71 &      0.027  &    \\
$\bar{p} (1710)$  &        1.71 &      0.027  &    \\
$\phi (1680)$  &        1.68 &      0.026  &    \\
$\bar{\Delta}^{0}(1950)$  &        1.93 &      0.024  &    \\
$\Delta^{+} (1950) $  &        1.93 &      0.024  &   99\% \\ \hline
$K^{*+} (1680) $  &       1.717 &      0.024  &    \\
$\bar{K}^{*0} (1680) $  &       1.717 &      0.024  &    \\
$\pi_1^{-} (1600) $  &       1.653 &      0.024  &    \\
$\pi_1^{0} (1600) $  &       1.653 &      0.024  &    \\
$\bar{\Sigma}^{-}(1915) $  &       1.915 &      0.023  &    \\
$\Sigma^{+} (1915)$  &       1.915 &      0.023  &    \\
$\bar{\Delta}^{0} (1920) $  &        1.92 &      0.023  &    \\
$\Delta^{+} (1920) $  &        1.92 &      0.023  &    \\
$\Xi^{0} (1820)$  &       1.823 &      0.023  &    \\
$\bar{\Xi}^{-} (1820)$  &       1.823 &      0.023  &    \\
$\bar{\Sigma}^{-}(1940)$  &        1.94 &      0.022  &    \\
$\Sigma^{+}(1940)$  &        1.94 &      0.022  &    \\
$\Lambda (1670)$  &        1.67 &      0.021  &    \\
$\bar{\Lambda} (1670)$  &        1.67 &      0.021  &    \\
$\bar{\Delta}^{-} (1910) $  &        1.91 &       0.02  &    \\
$\Delta^{++} (1910) $  &        1.91 &       0.02  &    \\
$\Sigma^{0} (1775)$  &       1.775 &       0.02  &    \\
$\bar{\Sigma}^{0} (1775)$  &       1.775 &       0.02  &    \\
$K_2^0 (1820)$  &       1.816 &      0.019  &    \\
$K_2^- (1820)$  &       1.816 &      0.019  &    \\
$\bar{\Delta}^{-}(1930)$  &        1.96 &      0.017  &    \\
$\Delta^{++} (1930) $  &        1.96 &      0.017  &    \\
\hline \hline
\end{tabular}
\caption{Resonance contribution list for $\pi^+$, for $T_\mathrm{conv}{\,=\,}120$\,MeV (continued).}
\label{table1_4}
\end{table}

\begin{table}
\begin{tabular}{c|c|c|c}
\hline \hline
name & mass (GeV) & total contribution (\%) &  \\ \hline
$f_2' (1525)$  &       1.525 &      0.015  &    \\
$\Sigma ^{0} (1660)$  &        1.66 &      0.014  &    \\
$\bar{\Sigma} ^{0} (1660)$  &        1.66 &      0.014  &    \\
$\bar{\Sigma} ^{-} (1750)$  &        1.75 &      0.014  &    \\
$\Sigma ^{+} (1750)$  &        1.75 &      0.014  &    \\
$\Sigma ^{0} (1750)$  &        1.75 &      0.014  &    \\
$\bar{\Sigma} ^{0} (1750)$  &        1.75 &      0.014  &    \\
$K_3^{*0} (1780) $  &       1.776 &      0.014  &    \\
$K_3^{*-} (1780) $  &       1.776 &      0.014  &    \\
$f_2(2010)$  &       2.011 &      0.012  &    \\
$\Delta^{-} (1600) $  &         1.6 &      0.012  &    \\
$\bar{\Delta}^{++} (1600) $  &         1.6 &      0.012  &    \\
$\Lambda (1890)$  &        1.89 &      0.011  &    \\
$\bar{\Lambda} (1890)$  &        1.89 &      0.011  &    \\
$\Delta^{0} (1920) $  &        1.92 &      0.011  &    \\
$\bar{\Delta}^{+} (1920) $  &        1.92 &      0.011  &    \\
$\bar{\Delta}^{0}(1910) $  &        1.91 &      0.011  &    \\
$\Delta^{+} (1910) $  &        1.91 &      0.011  &    \\
$\Delta^{0} (1950) $  &        1.93 &      0.011  &    \\
$\bar{\Delta}^{+} (1950) $  &        1.93 &      0.011  &    \\
$\Lambda (1820)$  &        1.82 &      0.011  &    \\
$\bar{\Lambda} (1820)$  &        1.82 &      0.011  &    \\
$\pi^{+} (1800) $  &       1.812 &       0.01  &    \\
$\Lambda (1800)$  &         1.8 &       0.01  &    \\
$\bar{\Lambda} (1800)$  &         1.8 &       0.01  &    \\
$\Sigma^{0}(1940) $  &        1.94 &      0.009  &    \\
$\bar{\Sigma}^{0}(1940) $  &        1.94 &      0.009  &    \\
$\Sigma ^{-} (1750)$  &        1.75 &      0.009  &    \\
$\bar{\Sigma} ^{+} (1750)$  &        1.75 &      0.009  &    \\
$\Sigma ^{0} (1915)$  &       1.915 &      0.009  &    \\
$\bar{\Sigma} ^{0} (1915)$  &       1.915 &      0.009  &    \\
$\bar{\Xi} ^{0} (1820)$  &       1.823 &      0.009  &    \\
$\Xi ^{-} (1820)$  &       1.823 &      0.009  &    \\
$\Xi^{0}(1950)$  &        1.95 &      0.008  &    \\
$\bar{\Xi}^{-}(1950) $  &        1.95 &      0.008  &    \\
$\Lambda (1810)$  &        1.81 &      0.008  &    \\
$\bar{\Lambda} (1810)$  &        1.81 &      0.008  &    \\
$K^{*-} (1680) $  &       1.717 &      0.007  &    \\
$K^{*0} (1680) $  &       1.717 &      0.007  &    \\
$\bar{\Sigma} ^{+} (1775)$  &       1.775 &      0.007  &    \\
$\Sigma ^{-} (1775)$  &       1.775 &      0.006  &    \\
$\bar{\Delta}^{0} (1930) $  &        1.96 &      0.006  &    \\
$\Delta^{+} (1930) $  &        1.96 &      0.006  &    \\
$\Omega(2250)$  &       2.252 &      0.006  &    \\
$\Delta^{0} (1910) $  &        1.91 &      0.005  &    \\
$\bar{\Delta}^{+} (1910) $  &        1.91 &      0.005  &    \\
$n (1650)$  &       1.655 &      0.005  &    \\
$\bar{p} (1650)$  &       1.655 &      0.005  &    \\
\hline \hline
\end{tabular}
\caption{Resonance contribution list for $\pi^+$, for $T_\mathrm{conv}{\,=\,}120$\,MeV (continued).}
\label{table1_5}
\end{table}

\begin{table}
\begin{tabular}{c|c|c|c}
\hline \hline
name & mass (GeV) & total contribution (\%) &  \\ \hline
$\pi^{-} (1800) $  &       1.812 &      0.004  &    \\
$\pi^{0} (1800) $  &       1.812 &      0.004  &    \\
$\phi_3 (1850)$  &       1.854 &      0.004  &    \\
$f_2(1950)$  &       1.945 &      0.003  &    \\
$\Delta^{-} (1920) $  &        1.92 &      0.003  &    \\
$\bar{\Delta}^{++} (1920) $  &        1.92 &      0.003  &    \\
$f_0 (1710)$  &       1.715 &      0.002  &    \\
$\Delta^{-} (1910) $  &        1.91 &      0.002  &    \\
$\bar{\Delta}^{++} (1910) $  &        1.91 &      0.002  &    \\
$\Sigma^{-}(1940) $  &        1.94 &      0.001  &    \\
$\bar{\Sigma}^{+} (1940)$  &        1.94 &      0.001  &    \\
$\Sigma ^{-}$ (1915) &       1.915 &          0  &    \\
$\bar{\Sigma} ^{+} (1915)$  &       1.915 &          0  &   100\%  \\
\hline \hline
\end{tabular}
\caption{Resonance contribution list for $\pi^+$, for $T_\mathrm{conv}{\,=\,}120$\,MeV (continued).}
\label{table1_6}
\end{table}


\begin{table}
\begin{tabular}{c|c|c|c}
\hline \hline
name & mass (GeV) & total contribution (\%) &  \\ \hline
$K^{*0}(892)$  &      0.8961 &     35.857  &    \\
$K^{*+}(892)$  &     0.89166 &      18.52  &    \\
$\phi(1020)$  &      1.0195 &     16.036  &   60\% \\ \hline
$K_1^{+} (1270) $  &       1.273 &      3.631  &    \\
$K_1^{0} (1270) $  &       1.273 &      3.287  &    \\
$a_0^{+}(980)$  &      0.9847 &      1.807  &    \\
$f_1 (1420)$  &      1.4263 &      1.754  &  80\%  \\ \hline
$K_2^{*0} (1430) $  &      1.4324 &       1.61  &    \\
$K_1^{+} (1400) $  &       1.402 &      1.446  &    \\
$K_2^{*+} (1430) $  &      1.4256 &      1.241  &    \\
$K^{*+} (1410) $  &       1.414 &      1.217  &    \\
$K_1^{0} (1400) $  &       1.402 &      1.127  &    \\
$K^{*0} (1410) $  &       1.414 &      1.033  &    \\
$a_0^{0}(980)$  &      0.9847 &      0.898  &    \\
$f_0(980)$  &      0.9741 &      0.869  &  90\%  \\ \hline
$\bar{\Lambda} (1520)$  &      1.5195 &      0.798  &    \\
$f_2' (1525)$  &       1.525 &      0.711  &    \\
$f_1 (1285)$  &      1.2818 &      0.493  &    \\
$\bar{\Xi} ^{-} (1690)$  &        1.69 &      0.462  &    \\
$\bar{\Sigma} ^{-} (1775)$  &       1.775 &      0.431  &    \\
$\eta (1475)$  &       1.476 &      0.405  &    \\
$K_0^{*0} (1430) $  &       1.412 &      0.295  &    \\
$\bar{\Xi} ^{-} (1820)$  &       1.823 &      0.288  &    \\
$a_2^{+} (1320) $  &      1.3183 &      0.274  &    \\
$\eta (1405)$  &      1.4103 &      0.265  &    \\
$\phi (1680)$  &        1.68 &      0.265  &  95\%  \\  \hline
$f_2(2010)$  &       2.011 &      0.259  &    \\
$\bar{\Sigma} ^{0} (1775)$  &       1.775 &      0.237  &    \\
$\bar{\Xi} ^{0} (1690)$  &        1.69 &       0.23  &    \\
$\bar{\Lambda} (1820)$  &        1.82 &      0.196  &    \\
$K_3^{*+} (1780) $  &       1.776 &      0.191  &    \\
$K_2^{+} (1770) $  &       1.773 &      0.189  &    \\
$\bar{\Sigma} ^{-} (1670)$  &        1.67 &      0.177  &    \\
$\bar{\Sigma} ^{-} (1750)$  &        1.75 &      0.171  &    \\
$K_2^{0} (1770) $  &       1.773 &      0.166  &    \\
$\bar{\Sigma} ^{-} (1660)$  &        1.66 &      0.165  &    \\
$\bar{\Lambda} (1600)$  &         1.6 &      0.161  &    \\
$\eta (1295)$  &       1.294 &      0.157  &    \\
$K_0^{*+} (1430) $  &       1.412 &      0.148  &    \\
$K_2^{+} (1820) $  &       1.816 &      0.145  &    \\
$\bar{\Lambda} (1690)$  &        1.69 &      0.144  &    \\
$K_3^{*0} (1780) $  &       1.776 &       0.14  &    \\
$a_2^{0} (1320) $  &      1.3183 &      0.137  &  98\%  \\ \hline
$K^{*0} (1680) $  &       1.717 &      0.117  &    \\
\hline \hline
\end{tabular}
\caption{Resonance contribution list for $K^+$, for $T_\mathrm{conv}{\,=\,}120$\,MeV.}
\label{table2_1}
\end{table}

\begin{table}
\begin{tabular}{c|c|c|c}
\hline \hline
name & mass (GeV) & total contribution (\%) &  \\ \hline
$\phi_3 (1850)$  &       1.854 &      0.114  &    \\
$f_2 (1270)$  &      1.2754 &      0.099  &    \\
$\bar{\Lambda} (1810)$  &        1.81 &      0.091  &    \\
$\bar{\Sigma} ^{0} (1670)$  &        1.67 &      0.089  &    \\
$\bar{\Lambda} (1890)$  &        1.89 &      0.087  &    \\
$\bar{\Sigma} ^{0} (1750)$  &        1.75 &      0.085  &    \\
$\bar{\Sigma} ^{0} (1660)$  &        1.66 &      0.082  &    \\
$K_2^{0} (1820) $  &       1.816 &      0.082  &    \\
$\bar{\Xi}^{-}(1950) $  &        1.95 &      0.082  &    \\
$\bar{\Lambda} (1670)$  &        1.67 &      0.082  &  99\%  \\ \hline
$K^{*+}(1680) $  &       1.717 &      0.078  &    \\
$\bar{\Lambda} (1800)$  &         1.8 &      0.076  &    \\
$p (1720)$  &        1.72 &      0.072  &    \\
$p (1710)$  &        1.71 &      0.068  &    \\
$\eta_2 (1645)$  &       1.617 &       0.06  &    \\
$\bar{\Xi} ^{0} (1820)$  &       1.823 &      0.052  &    \\
$a_0^{+} (1450) $  &       1.474 &      0.052  &    \\
$\bar{\Sigma}^{+} (1940)$  &        1.94 &      0.047  &    \\
$\bar{\Sigma}^{0}(1940) $  &        1.94 &      0.045  &    \\
$\bar{\Omega}(2250)$  &       2.252 &      0.044  &    \\
$\bar{\Sigma}^{+} (1775)$  &       1.775 &      0.043  &    \\
$\bar{\Sigma}^{-}(1940)$  &        1.94 &      0.043  &    \\
$\bar{\Sigma}^{-}(1915) $  &       1.915 &      0.036  &    \\
$p (1650)$  &       1.655 &      0.032  &    \\
$a_0^{0} (1450) $  &       1.474 &      0.026  &    \\
$f_0 (1710)$  &       1.715 &      0.024  &    \\
$\rho_3^{+} (1690) $  &      1.6888 &      0.023  &    \\
$\bar{\Sigma} ^{0} (1915)$  &       1.915 &      0.018  &    \\
$\bar{\Lambda} (1830)$  &        1.83 &      0.017  &    \\
$\rho_3^{0} (1690) $  &      1.6888 &      0.016  &    \\
$\pi_2^{+} (1670) $  &      1.6724 &      0.015  &    \\
$\pi_2^{0} (1670)  $  &      1.6724 &      0.013  &    \\
$f_0 (1500)$  &       1.507 &      0.012  &    \\
$\pi_2^{-} (1670)  $  &      1.6724 &       0.01  &    \\
$f_0 (1370)$  &         1.4 &       0.01  &    \\
$\rho_3^{-} (1690) $  &      1.6888 &      0.008  &    \\
$\Delta^{++}(1920) $  &        1.92 &      0.004  &    \\
$\Delta^{+} (1920) $  &        1.92 &      0.003  &    \\
$\Delta^{++} (1950) $  &        1.93 &      0.003  &    \\
$\Delta^{+} (1950) $  &        1.93 &      0.002  &    \\
$\bar{K}_1^{0} (1400) $  &       1.402 &      0.002  &    \\
$K_1^{-} (1400) $  &       1.402 &      0.002  &    \\
$\Delta^{0} (1920) $  &        1.92 &      0.001  &    \\
$f_2(1950)$  &       1.945 &      0.001  &    \\
$\Delta^{0} (1950) $  &        1.93 &      0.001  &    \\
$\bar{K}_2^{0} (1820)$  &       1.816 &          0  &    \\
$K_2^{-} (1820) $  &       1.816 &          0  &   100\%  \\\hline \hline
\end{tabular}
\caption{Resonance contribution list for $K^+$, for $T_\mathrm{conv}{\,=\,}120$\,MeV (continued).}
\label{table2_2}
\end{table}

\begin{table}
\begin{tabular}{c|c|c|c}
\hline \hline
name & mass (GeV) & total contribution (\%) &  \\ \hline
$\Delta^{++} (1232) $  &       1.232 &     29.842  &    \\
$\Delta^{+} (1232) $  &       1.232 &     19.816  &    \\
$\Delta^{0} (1232) $  &       1.232 &      9.813  &    \\
$\Delta^{++} (1600) $  &         1.6 &      2.787  &  60\%  \\ \hline
$n (1520)$  &        1.52 &      2.487  &    \\
$p (1520)$  &        1.52 &      2.169  &    \\
$\Delta^{+} (1600) $  &         1.6 &      2.049  &    \\
$p (1440)$  &        1.44 &      2.034  &    \\
$n (1440)$  &        1.44 &      1.943  &    \\
$p (1535)$  &       1.535 &      1.452  &    \\
$\Delta^{++} (1700) $  &         1.7 &      1.386  &    \\
$\Lambda (1520)$  &      1.5195 &      1.365  &    \\
$p (1675)$  &       1.675 &      1.347  &    \\
$\Delta^{0} (1600) $  &         1.6 &      1.314  &    \\
$p (1700)$  &         1.7 &      1.228  &  80\%   \\ \hline
$\Delta^{++} (1620) $  &        1.63 &      1.135  &    \\
$n (1680)$  &       1.685 &      1.003  &    \\
$\Delta^{+} (1700) $  &         1.7 &      0.987  &    \\
$n (1675)$  &       1.675 &      0.975  &    \\
$p (1680)$  &       1.685 &      0.917  &    \\
$n (1535)$  &       1.535 &      0.793  &    \\
$\Delta^{+} (1620) $  &        1.63 &      0.785  &    \\
$n (1720)$  &        1.72 &      0.752  &    \\
$\Sigma ^{+} (1775)$  &       1.775 &      0.738  &    \\
$\Delta^{-} (1600) $  &         1.6 &      0.574  &    \\
$\Delta^{0} (1700) $  &         1.7 &       0.54  &   90\% \\ \hline
$\Delta^{++} (1905) $  &        1.89 &      0.536  &    \\
$p (1720)$  &        1.72 &      0.487  &    \\
$p (1710)$  &        1.71 &      0.456  &    \\
$\Delta^{0} (1620) $  &        1.63 &      0.433  &    \\
$n (1650)$  &       1.655 &      0.429  &    \\
$\Delta^{++} (1950)$  &        1.93 &      0.411  &    \\
$\Sigma ^{0} (1775)$  &       1.775 &      0.405  &    \\
$\Delta ^{+} (1905)$  &        1.89 &      0.366  &    \\
$n (1700)$  &         1.7 &      0.342  &    \\
$\Lambda (1820)$  &        1.82 &      0.334  &    \\
$\Sigma ^{+} (1670)$  &        1.67 &      0.302  &    \\
$p (1650)$  &       1.655 &       0.29  &  95\%  \\ \hline
$\Sigma ^{+} (1750)$  &        1.75 &      0.289  &    \\
$\Sigma^{+} (1660)$  &        1.66 &      0.282  &    \\
$\Delta^{+} (1950) $  &        1.93 &      0.281  &    \\
$\Lambda (1600)$  &         1.6 &      0.275  &    \\
$\Lambda (1690)$  &        1.69 &      0.247  &    \\
$\Delta ^{++}(1920)$  &        1.92 &      0.227  &    \\
$n (1710)$  &        1.71 &      0.203  &    \\
$\Delta^{++} (1930) $  &        1.96 &      0.196  &    \\
$\Delta^{0} (1905) $  &        1.89 &      0.189  &    \\
$\Delta^{+} (1920)$  &        1.92 &      0.182  &    \\
\hline \hline
\end{tabular}
\caption{Resonance contribution list for $p$, for $T_\mathrm{conv}{\,=\,}120$\,MeV.}
\label{table3_1}
\end{table}

\begin{table}
\begin{tabular}{c|c|c|c}
\hline \hline
name & mass (GeV) & total contribution (\%) &  \\ \hline
$\Lambda (1810)$  &        1.81 &      0.156  &    \\
$\Delta^{0} (1950)$  &        1.93 &      0.153  &    \\
$\Sigma ^{0} (1670)$  &        1.67 &      0.151  &    \\
$\Lambda (1890)$  &        1.89 &       0.15  &  98\%  \\ \hline
$\Sigma ^{0} (1750)$  &        1.75 &      0.145  &    \\
$\Sigma ^{0} (1660)$  &        1.66 &      0.141  &    \\
$\Lambda (1670)$  &        1.67 &      0.139  &    \\
$\Sigma^{+}(1940)$  &        1.94 &      0.138  &    \\
$\Delta^{0} (1920) $  &        1.92 &      0.137  &    \\
$\Delta^{-} (1700) $  &         1.7 &      0.136  &    \\
$\Delta^{+} (1930) $  &        1.96 &       0.13  &    \\
$\Lambda (1800)$  &         1.8 &       0.13  &  99\%  \\ \hline
$\Delta^{++} (1910) $  &        1.91 &      0.121  &    \\
$\Delta^{+} (1910) $  &        1.91 &      0.096  &    \\
$\Delta^{-} (1920) $  &        1.92 &      0.092  &    \\
$\Delta^{-} (1620) $  &        1.63 &      0.081  &    \\
$\Sigma^{0}(1940) $  &        1.94 &      0.077  &    \\
$\Sigma ^{-} (1775)$  &       1.775 &      0.074  &    \\
$\Delta^{0} (1910) $  &        1.91 &      0.071  &    \\
$\Delta^{0} (1930) $  &        1.96 &      0.065  &    \\
$\Sigma ^{+} (1915)$  &       1.915 &      0.062  &    \\
$\Delta^{-} (1910) $  &        1.91 &      0.046  &    \\
$\Sigma ^{0} (1915)$  &       1.915 &      0.031  &    \\
$\Lambda (1830)$  &        1.83 &      0.029  &    \\
$\Delta^{-}(1950) $  &        1.93 &      0.023  &    \\
$\Delta^{-} (1905) $  &        1.89 &      0.016  &    \\
$\Sigma^{-}(1940) $  &        1.94 &      0.015  &  100\%  \\
\hline \hline
\end{tabular}
\caption{Resonance contribution list for $p$, for $T_\mathrm{conv}{\,=\,}120$\,MeV (continued).}
\label{table3_2}
\end{table}

\begin{table}
\begin{tabular}{c|c|c|c}
\hline \hline
name & mass (GeV) & total contribution (\%) &  \\ \hline
$\Sigma ^{0}$  &      1.1926 &     24.775  &    \\
$\Sigma ^{+}(1385) $  &      1.3828 &     17.893  &    \\
$\Sigma^{-} (1385) $  &      1.3872 &     17.346  &   60\% \\ \hline
$\Sigma^{0} (1385) $  &      1.3837 &     16.555  &    \\
$\Lambda (1405)$  &      1.4065 &      3.103  &    \\
$\Lambda (1520)$  &      1.5195 &       2.18  &   80\% \\ \hline
$\Sigma ^{-} (1670)$  &        1.67 &      1.304  &    \\
$\Sigma ^{+} (1670)$  &        1.67 &      1.297  &    \\
$\Lambda (1690)$  &        1.69 &      1.178  &    \\
$\Xi ^{-} (1690)$  &        1.69 &      1.132  &    \\
$\Xi ^{0} (1690)$  &        1.69 &      1.128  &    \\
$\Sigma ^{+} (1775)$  &       1.775 &      0.771  &    \\
$\Sigma ^{-} (1775)$  &       1.775 &      0.769  &    \\
$\Lambda (1830)$  &        1.83 &      0.754  &  90\%  \\ \hline
$\Sigma ^{0} (1775)$  &       1.775 &      0.749  &    \\
$\Sigma ^{-} (1660)$  &        1.66 &      0.708  &    \\
$\Sigma^{+} (1660)$  &        1.66 &      0.707  &    \\
$\Xi ^{0} (1820)$  &       1.823 &      0.707  &    \\
$\Xi ^{-} (1820)$  &       1.823 &      0.705  &    \\
$\Sigma ^{0} (1750)$  &        1.75 &      0.619  &    \\
$\Lambda (1670)$  &        1.67 &      0.607  &    \\
$\Lambda (1600)$  &         1.6 &      0.486  &  95\%  \\ \hline
$\Sigma^{0} (1670)$  &        1.67 &      0.485  &    \\
$\Sigma^{0} (1660)$  &        1.66 &      0.471  &    \\
$\Sigma^{-} (1915)$  &       1.915 &      0.382  &    \\
$\Sigma ^{+} (1915)$  &       1.915 &      0.381  &    \\
$\Sigma ^{0} (1915)$  &       1.915 &      0.264  &    \\
$\Lambda (1820)$  &        1.82 &      0.216  &    \\
$\Xi^{0}(1950)$  &        1.95 &      0.202  &    \\
$\Xi ^{-} (1950)$  &        1.95 &      0.201  &  98\%  \\ \hline
$\Sigma^- (1940)$  &        1.94 &      0.189  &    \\
$\Sigma^+ (1940)$  &        1.94 &      0.188  &    \\
$n (1720)$  &        1.72 &      0.176  &    \\
$p (1720)$  &        1.72 &      0.176  &    \\
$\Lambda (1890)$  &        1.89 &      0.174  &    \\
$n (1710)$  &        1.71 &      0.167  &   99\%  \\ \hline
$p (1710)$  &        1.71 &      0.167  &    \\
$\Sigma^0 (1940)$  &        1.94 &      0.155  &    \\
$\Lambda (1800)$  &         1.8 &      0.155  &    \\
$n (1650)$  &       1.655 &      0.078  &    \\
$p (1650)$  &       1.655 &      0.078  &    \\
$\Sigma ^{-} (1750)$  &        1.75 &      0.076  &    \\
$\Sigma ^{+} (1750)$  &        1.75 &      0.075  &    \\
$\Lambda (1810)$  &        1.81 &       0.05  &    \\
$\Delta^{0} (1920) $  &        1.92 &      0.007  &    \\
$\Delta^{+} (1920) $  &        1.92 &      0.007  &    \\
$\Delta^{0} (1950)$  &        1.93 &      0.004  &    \\
$\Delta^{+} (1950) $  &        1.93 &      0.004  &  100\%  \\
\hline \hline
\end{tabular}
\caption{Resonance contribution list for $\Lambda$, for $T_\mathrm{conv}{\,=\,}120$\,MeV.}
\label{table4}
\end{table}

\begin{table}
\begin{tabular}{c|c|c|c}
\hline \hline
name & mass (GeV) & total contribution (\%) &  \\ \hline
$\Lambda (1405)$  &      1.4065 &     25.159  &    \\
$\Lambda (1520)$  &      1.5195 &     10.121  &    \\
$\Sigma^{+} (1385) $  &      1.3828 &      9.288  &    \\
$\Sigma^{0} (1385) $  &      1.3837 &      9.179  &    \\
$\Sigma ^{0} (1670)$  &        1.67 &      6.648  &  60\%  \\ \hline
$\Sigma ^{+} (1670)$  &        1.67 &      6.618  &    \\
$\Xi ^{0} (1690)$  &        1.69 &      4.578  &    \\
$\Sigma ^{+} (1750)$  &        1.75 &      4.425  &    \\
$\Lambda (1600)$  &         1.6 &      3.969  &    \\
$\Lambda (1690)$  &        1.69 &      3.845  &   80\% \\ \hline
$\Lambda (1830)$  &        1.83 &      2.243  &    \\
$\Lambda (1670)$  &        1.67 &      1.926  &    \\
$\Sigma^{+} (1660)$  &        1.66 &      1.916  &    \\
$\Sigma ^{0} (1660)$  &        1.66 &      1.916  &   90\%  \\ \hline
$\Xi ^{0} (1820)$  &       1.823 &      1.045  &    \\
$\Sigma ^{+} (1915)$  &       1.915 &      0.974  &    \\
$\Sigma ^{0} (1915)$  &       1.915 &      0.968  &    \\
$\Sigma ^{+} (1775)$  &       1.775 &      0.859  &  95\%  \\ \hline
$\Sigma ^{0} (1775)$  &       1.775 &      0.805  &    \\
$\Lambda (1820)$  &        1.82 &      0.791  &    \\
$\Sigma ^{-} (1775)$  &       1.775 &        0.6  &    \\
$\Sigma^{+}(1940)$  &        1.94 &      0.405  &  98\%  \\ \hline
$\Lambda (1810)$  &        1.81 &      0.402  &    \\
$\Sigma^{0}(1940) $  &        1.94 &      0.387  &   99\%  \\ \hline
$\Lambda (1800)$  &         1.8 &      0.377  &    \\
$\Lambda (1890)$  &        1.89 &      0.231  &    \\
$\Sigma^{-}(1940) $  &        1.94 &      0.101  &    \\
$\Delta^{++}(1920) $  &        1.92 &       0.08  &    \\
$\Delta^{++} (1950) $  &        1.93 &      0.051  &    \\
$\Delta^{+} (1920) $  &        1.92 &      0.027  &    \\
$\Sigma ^{-} (1750)$  &        1.75 &       0.02  &    \\
$\Sigma ^{0} (1750)$  &        1.75 &       0.02  &    \\
$\Delta^{+} (1950) $  &        1.93 &      0.015  &    \\
$\Sigma ^{-} (1915) $  &       1.915 &      0.007  &   100\% \\
\hline \hline
\end{tabular}
\caption{Resonance contribution list for $\Sigma^+$, for $T_\mathrm{conv}{\,=\,}120$\,MeV.}
\label{table2}
\end{table}

\begin{table}
\begin{tabular}{c|c|c|c}
\hline \hline
name & mass (GeV) & total contribution (\%) &  \\ \hline
$\Xi^{0} (1530)$  &      1.5318 &     62.049  &   60\% \\  \hline
$\Xi ^{-} (1530)$  &       1.535 &     30.103  &    90\% \\ \hline
$\Xi ^{0} (1820)$  &       1.823 &       2.42  &    \\
$\Xi ^{-} (1820)$  &       1.823 &      2.334  &    95\% \\ \hline
$\Xi^{0} (1950)$  &        1.95 &      1.427  &     98\% \\ \hline
$\Omega(2250)$  &       2.252 &      0.957  &  99\%  \\ \hline
$\Xi ^{-} (1950)$  &        1.95 &       0.71  &    100\% \\
\hline \hline
\end{tabular}
\caption{Resonance contribution list for $\Xi^-$, for $T_\mathrm{conv}{\,=\,}120$\,MeV.}
\label{table6}
\end{table}

\end{document}